\PassOptionsToPackage{protrusion, expansion}{microtype} 
\PassOptionsToPackage{style=ieee, citestyle=numeric, sortcites=true, backend=bibtex}{biblatex}

\documentclass[conference, nofonttune]{IEEEtran}\def\UseIEEETemplate{1}  
\IEEEoverridecommandlockouts

\usepackage{graphicx}
\usepackage{xcolor}
\definecolor{B}    {HTML}{2b66d3}   
\definecolor{B2}   {HTML}{003399}   
\definecolor{Bv}   {HTML}{0000EB}   
\definecolor{R}    {HTML}{c9171e}
\definecolor{R2}   {HTML}{d7003a}
\definecolor{INK}  {HTML}{595857}
\definecolor{Y}    {HTML}{f1c40f}
\definecolor{G}    {HTML}{009a00}
\definecolor{GRAY} {HTML}{808080}
\definecolor{MAUVE}{HTML}{9400D1}
\usepackage{amsmath,amsfonts}    
\usepackage{mathtools}
\usepackage{bm}
\usepackage{amsthm}

\usepackage{array} 
\usepackage{booktabs}
\usepackage{colortbl}
\usepackage{adjustbox}
\usepackage{multirow}

\usepackage{url, hyperref}                  
\hypersetup{
    pdfborder={0 0 0}}

\usepackage{enumitem}                       
\usepackage{lipsum}                         
\usepackage[plain]{algorithm}       		
\floatstyle{plaintop}
\usepackage{algpseudocode}
\algrenewcommand{\alglinenumber}[1]{{\scriptsize\bfseries\ttfamily\color{R}#1}}
\floatstyle{plaintop}
\restylefloat{algorithm}

\usepackage{xpatch}
\makeatletter
\xpatchcmd{\algorithmic}{\ALG@tlm\z@}{\ALG@tlm\z@\leftmargin 10pt}{}{}
\makeatother

\usepackage{listings}						
\usepackage{caption, subcaption}            
\usepackage{setspace}                       

\usepackage{soul}
\ifx\UseACMTemplate\undefined
\else
    \newcommand{\SEC}{\textcolor{black}{\S}}
    \newcommand{\FIG}{\textcolor{black}{Figure}}
    \newcommand{\TAB}{\textcolor{black}{Table}}
\fi

\ifx\UseIEEETemplate\undefined
\else
    \newcommand{\SEC}{\textcolor{black}{\S}}
    \newcommand{\FIG}{\textcolor{black}{Fig.}}
    \newcommand{\TAB}{\textcolor{black}{TABLE}}
\fi

\usepackage{comment}

\newcommand{\BOLD}{\fontfamily{ugq}\selectfont}

\newcommand{\TEXT}[1]{\text{\ttfamily[#1]}}
\newcommand{\TT}[1]{\text{\ttfamily#1}}

\renewcommand{\TEXT}[1]{{[#1]}}
\renewcommand{\TT}[1]{{#1}}
\usepackage{tikz}
\usetikzlibrary{positioning}        
\usetikzlibrary{shapes.multipart}   
\usetikzlibrary{shapes.geometric}   
\usetikzlibrary{arrows}
\usetikzlibrary{arrows.meta}        
\usetikzlibrary{backgrounds}
\usetikzlibrary{patterns}

\usepackage{wrapfig}

\newcommand{\TABLETITLE}{\BOLD\color{Bv}}
\newcommand{\TABLECAPTION}{\color{gray}\scriptsize}

\newcommand{\PartialSum}[2]{%
	\left.p\Sigma\left({\color{black}#1}\right)\right\vert_{#2}%
}
\newcommand{\QuantAddend}[2]{%
+q'_{#1,#2}%
}

\newcommand{%
	\tableHeadBold}[1]{\multicolumn{1}{c}{\sffamily\bfseries\begin{tabular}{@{}c@{}}#1\end{tabular}}%
}

\newcommand{%
	\notAvailable}{{\setlength{\fboxsep}{1pt}\fbox{$\times$}}%
}
\usepackage{biblatex}
\newcommand{\cusz}{\textsc{cuSZ}}       
\newcommand{\cuszx}{\textsc{cuSZ+}}   

\newcommand{\avgbits}{\langle b\rangle}

\newcommand{\quantcode}{quant-code}

\usepackage{titlesec}

\renewcommand\thesubsubsection{\Alph{subsection}.\arabic{subsubsection}}
\renewcommand\theparagraph{\underline{\Alph{subsection}.\arabic{subsubsection}.\alph{paragraph}}}

\titleformat{\subsubsection}[runin]
{\normalfont\normalsize\itshape}{\thesubsubsection)}{.6em}{}[\ \ ]
\titlespacing*{\subsubsection}
{0pt}{.3ex plus .1ex minus .1ex}{.3ex plus .1ex minus .1ex}

\titleformat{\paragraph}[runin]
{\normalfont\normalsize\itshape}{\theparagraph}{.5em}{}[\ \ ]
\titlespacing*{\paragraph}
{0pt}{.25ex plus 1ex minus .2ex}{.25ex plus .2ex}

\usepackage[scaled=.98]{inconsolata}
\usepackage[scale=.92]{tgheros}
\usepackage[normalem]{ulem}

\hypersetup{
	pdftitle={Optimizing Error-Bounded Lossy Compression for Scientific Data on GPUs},
	pdfauthor={Tian et al.},
	pdfsubject={cuSZ+, camera-ready},
	pdfkeywords={GPU; compression; lossy; error bound; HPC; scientific data}
}

\begin{document}
\title{Optimizing Error-Bounded Lossy Compression for Scientific Data on GPUs}

\newcommand{\FirstAuthorMark}{$^\star$}

\author{Jiannan Tian\FirstAuthorMark,
Sheng Di\IEEEauthorrefmark{2},
Xiaodong Yu\IEEEauthorrefmark{2},
Cody Rivera\IEEEauthorrefmark{4},
Kai Zhao\IEEEauthorrefmark{3},
Sian Jin\FirstAuthorMark,\\
Yunhe Feng\IEEEauthorrefmark{5},
Xin Liang\IEEEauthorrefmark{6},
Dingwen Tao\FirstAuthorMark,
Franck Cappello\IEEEauthorrefmark{2}\thanks{Corresponding author: Dingwen Tao (\url{dingwen.tao@wsu.edu}), School of EECS, Washington State University, Pullman, WA 99164, USA.}\\
\IEEEauthorblockA{\FirstAuthorMark Washington State University, Pullman, WA, USA}
\IEEEauthorblockA{\IEEEauthorrefmark{2}Argonne National Laboratory, Lemont, IL, USA}
\IEEEauthorblockA{\IEEEauthorrefmark{3}University of California, Riverside, Riverside, CA, USA}
\IEEEauthorblockA{\IEEEauthorrefmark{4}University of Alabama, Tuscaloosa, AL, USA}
\IEEEauthorblockA{\IEEEauthorrefmark{5}University of Washington, Seattle, WA, USA}
\IEEEauthorblockA{\IEEEauthorrefmark{6}Missouri University of Science and Technology, Rolla, MO, USA}
}

\newcommand{\MARK}[1]{\textcolor{B}{\sffamily #1}}
\newcommand{\cody}[1]{\textcolor{G}{#1}}

\maketitle

\thispagestyle{plain}
\pagestyle{plain}

\begin{abstract}
	Error-bounded lossy compression is a critical technique for significantly reducing scientific data volumes. With
	ever-emerging heterogeneous high-performance computing (HPC) architecture, GPU-accelerated error-bounded compressors (such as {\cusz} and cuZFP) have been developed. However, they suffer from either low performance or low compression ratios. To this end, we propose {\cuszx} to target both high compression ratios and throughputs. We identify that data sparsity and data smoothness are key factors for high compression throughputs. Our key contributions in this work are fourfold: (1) We propose an efficient compression workflow to adaptively perform run-length encoding and/or variable-length encoding. (2) We derive Lorenzo reconstruction in decompression as multidimensional partial-sum computation and propose a fine-grained Lorenzo reconstruction algorithm for GPU architectures. (3) We carefully optimize each of {\cusz} kernels by leveraging state-of-the-art CUDA parallel primitives. (4) We evaluate {\cuszx} using seven real-world HPC application datasets on V100 and A100 GPUs. Experiments show {\cuszx} improves the compression throughputs and ratios by up to 18.4$\times$ and 5.3$\times$, respectively, over {\cusz} on the tested datasets.
\end{abstract}

\section{Introduction}
\label{sec:intro}

Large-scale scientific applications for advanced instruments produce vast volumes of data every day for post hoc analysis.
For instance, Hardware/Hybrid Accelerated Cosmology Code (HACC)~\cite{hacc,miraio} may produce petabytes of data in hundreds of snapshots when simulating 1 trillion particles.
It could be very inefficient to store such a large amount of data, especially in situations with relatively low I/O bandwidth on the parallel file system (PFS) \cite{liang2018error,xincluster18}.

Data reduction is becoming an effective method to resolve the big-data issue for scientific research.
Although traditional lossless data reduction methods such as data deduplication and lossless compression can guarantee no information loss, they suffer from limited compression ratios on scientific datasets.
Specifically, deduplication usually reduces the scientific data size by only 20\% to 30\%~\cite{meister2012study}, and lossless compression achieves a compression ratio of up to $\sim$2:1~\cite{son2014data}.
The 2:1 is far lower than scientists' desired compression ratios (e.g., 10:1~\cite{use-case-Franck}).

Error-bounded lossy compressors have been developed for years to address the issue of low compression ratio for scientific data:
they can not only get very high compression ratios (such as over 100$\times$) \cite{sz16,sz17,zfp,liang2018error}, but strictly control the data distortion regarding the user-set error bound.
Notably, a qualified lossy compressor designed for scientific data reduction should address three primary concerns simultaneously: (1) high fidelity preservation, (2) high compression ratio, and (3) high throughput.
Most of the existing error-bounded lossy compressors (such as SZ \cite{sz16,sz17}, FPZIP \cite{fpzip}, ZFP \cite{zfp}), however, are mainly designed for CPU architectures, which cannot adapt to the high throughput requirement.
For example, LCLS-II laser \cite{lcls}, X-ray imaging data generated on advanced instruments, can result in a data acquisition rate at 250 GB/s \cite{use-case-Franck}.
As such, high compression throughput is critical for storing a tremendous amount of data efficiently for scientific projects.

Currently, several GPU-based error-controlled lossy compressors (such as {\cusz} \cite{cusz2020} and cuZFP
\cite{cuZFP}) have been developed, but they suffer from either sub-optimal compression throughputs or low compression ratios.
For instance, {\cusz} can achieve much higher compression ratios than cuZFP. Still, its performance is substantially
limited by the Huffman encoding and dictionary encoding stages when compared with the up-to-date work \cite{tian2021revisiting}.
However, the high compression ratios of SZ/{\cusz} significantly depend on Huffman encoding and dictionary encoding because the output of the prediction-and-quantization step in SZ/{\cusz} is often composed of many repeated symbols.

In this paper, we propose an efficient compression framework (called {\cuszx}) based on the {\cusz} framework, which can get both high compression ratios and high throughputs on GPUs.
The notation ``+'' in {\cuszx} indicates that this new compression method is specifically optimized for high performance in compression and decompression on the latest GPU architecture (i.e., NVIDIA's Ampere architecture).

It is challenging to develop an efficient GPU-based error-bounded lossy compressor that can achieve high compression ratios and high throughputs at the same time.
On the one hand, to develop efficient GPU code, one must maximize the parallelism from GPU threads.
Moreover, the architecture/characteristics of GPU accelerators (such as coherence, divergence issues, bank conflicts, use of shared memory, use of registers) must be coped with very carefully to get the optimal performance.
On the other hand, state-of-the-art error-bounded lossy compressors (such as SZ \cite{sz16,sz17,liang2018error}) often rely on Huffman encoding and dictionary encoding, which are procedures that contain substantial data dependencies.
These dependencies make them very hard to parallelize on GPUs efficiently.
For example, a Huffman tree must be built based on a code-frequency histogram before performing Huffman encoding, which has a significant data dependency inside.
The {\cusz} code just used one single GPU thread to do this work for simplicity.
Moreover, it is fairly non-trivial to design an efficient parallel code for the dictionary encoding because of the intrinsic dependency in its repeated sequence search.
	{\cusz} leaves this part to CPU, which may suffer from significant overhead.
Our key contributions proposed particularly in {\cuszx} are summarized as follows.
\begin{itemize}[noitemsep, topsep=2pt, leftmargin=1.3em]
	\item We design an adaptive compression workflow to perform run-length encoding and/or variable-length encoding
	      (i.e., Huffman encoding) on GPUs. We exploit a sufficient condition to determine when the run-length encoding should be applied for improving compression ratio, i.e., when the average Huffman bit-length is no greater than 1.09.
	\item We identify and prove that the first-order Lorenzo reconstruction in decompression is equivalent to a multidimensional partial-sum computation. We propose a fine-grained Lorenzo reconstruction algorithm based on a multidimensional partial-sum and a modified quantization scheme. Such a design can fully parallelize the decompression operation with workload tuning of GPU thread, improving the overall decompression throughput significantly.
	\item We develop some optimization strategies to boost compression performance and scalability. For instance,
	      we carefully optimize each kernel in compression considering CUDA architecture (e.g., reducing global memory accesses) to improve the compression throughput. We also leverage the state-of-the-art \texttt{NVIDIA::cub}
	      parallel primitives \cite{repo-NVIDIA-cub} to enhance the decompression scalability and throughput.
	\item We evaluate {\cuszx} with seven real-world HPC application datasets from public \emph{Scientific Data Reduction Benchmarks}~\cite{repo-sdrbench} on two state-of-the-art GPUs--V100 and A100. Experiments show that {\cuszx} improves the compression throughputs and ratios by up to 18.4$\times$ and 5.3$\times$, respectively, over {\cusz} on the tested datasets.
	\item We conclude that with the advancement of GPU architecture,  {\cuszx} can benefit more from the improvement of memory bandwidth than that of peak FLOPS and provide valuable insights for software and application R\&D toward the exascale computing era.
\end{itemize}

\section{Background and Research Motivation}
\label{sec:sz-bg}

In this section, we introduce the background of {\cusz} (the CUDA version of SZ) \cite{cusz2020} and our research motivation.

\subsection{Background of {\cusz}}

Unlike CPU-based SZ that has only four steps (prediction, quantization, Huffman encoding, and dictionary encoding), {\cusz} involves nine steps to adapt to the GPU architecture. Specifically, Step\nobreakdash-1 splits the whole dataset into multiple blocks, each of which will be compressed independently.
This design favors coarse-grained decompression.
Upon splitting blocks, \cusz's compression adopts a dual-quantization scheme (including prequantization%
\footnote{All data items are quantized based on their original values before the data prediction step.}%
, prediction, and postquantization), which can entirely remove the data dependency for the Lorenzo prediction.
Then, Step\nobreakdash-5 adopts parallel histograming to compute the frequencies of the {\quantcode}s.
Step\nobreakdash-6 builds a canonical Huffman codebook \cite{cusz2020} based on the histogram/frequency vector.
Step\nobreakdash-7 performs the Huffman encoding over the {\quantcode}s.
Step\nobreakdash-8 concatenates all the Huffman codes (called deflating) on GPUs, which feeds a dictionary encoder (Zstd \cite{zstd}) for further compression on CPUs in Step\nobreakdash-9.
The decompression is the reversed procedure of the compression.
We refer readers to the {\cusz} paper \cite{cusz2020} for more details.

For compression, Step\nobreakdash-6 and \nobreakdash-9 are the main bottlenecks because Step\nobreakdash-6 has to be
executed sequentially with a single GPU thread and designing an efficient multi-thread GPU algorithm for dictionary
encoding is non-trivial.

For decompression, the first step (i.e., the reversed dual-quantization) is the main bottleneck since the decompression cannot use the massive parallelism as the prequantization step does.
Specifically, in the decompression stage, the data values must be reconstructed one by one, according to the Lorenzo predictor%
\footnote{Lorenzo predictor predicts the data values based on a high-order data
	approximation formula: e.g., $X_{[j,i]} \approx X_{[j-1,i]} + X_{[j,i-1]} - X_{[j-1,i-1]}$ for 2D dataset, where $X_{[j,i]}$ refers to the value of the data item $[j,i]$ in the dataset.}.
To address this issue, {\cusz} adopts a coarse-grained parallel method instead: letting one GPU thread handle one independent data block in parallel.
However, such a design suffers from low performance due to the underuse of massive parallelism on GPUs.
In addition to Step\nobreakdash-1, Step\nobreakdash-9 is another significant bottleneck because the dictionary decoding is also very hard to parallelize on GPUs because of its intrinsic data dependency.

\subsection{Research Motivations}

\subsubsection{Limitation of {\cusz}'s Compression Ratio}

Full-fledged CPU-based compressors may utilize multifold techniques to boost the compression ratio: pattern-finding (e.g., BWT, LZ77), dictionary (e.g., LZ77), and variable-length encoding (``VLE'').
An exemplary combination is DEFLATE (LZ77 and VLE), whose famous implementation is \texttt{gzip} (used by CPU-SZ).
In comparison, {\cusz}, the GPU-based lossy compressor, only leverages Huffman coding to compress prediction-error correction codes (i.e., {\quantcode}s).
Even though Huffman coding is a VLE that is optimal in bit-length (i.e., with minimal discrepancy from the entropy), no less than one bit represents a data element. Therefore, {\cusz} can achieve up to 32$\times$ or 64$\times$ of compression ratios. {\cusz} disregards the repeated pattern that may exist in the symbol sequence. Hence, there is a gap between {\cusz} and CPU-SZ in compression ratio due to the lack of pattern-finding.

\begin{table}[ht]
	\centering\footnotesize\sffamily
\newcommand{\corcmidrule}[1][.75pt]{
     \\[\dimexpr-\normalbaselineskip-\belowrulesep-\aboverulesep-#1\relax]%
}

\newcommand{\BARqhgGRAY}[1]{%
     \scriptsize #1$\times$%
}
\newcommand{\BARqhg}[1]{%
     \scriptsize\color{Bv}#1$\times$%
}

\centering\footnotesize\ttfamily
\resizebox{.8\linewidth}{!}{
     \begin{tabular}{@{} >{\sffamily}r >{\color{gray}}rrr >{\color{gray}}rrr @{}}
               &
          \multicolumn{3}{c}{\TABLETITLE HACC}
               &
          \multicolumn{3}{c}{\TABLETITLE Hurricane}
          \\
               &
          \bfseries\textit{qg}
               &
          \bfseries\textit{qh}
               &
          \bfseries\textit{qhg}
               &
          \bfseries\textit{qg}
               &
          \bfseries\textit{qh}
               &
          \bfseries\textit{qhg}
          \\[1ex]
          1e-2 & 22.72            & 20.33        & 31.02        & 43.67            & 24.80        & 58.76        \\[-.5ex]
               & \BARqhgGRAY{1.1} & \BARqhg{1.0} & \BARqhg{1.5} & \BARqhgGRAY{1.8} & \BARqhg{1.0} & \BARqhg{2.4} \\
          1e-3 & 7.58             & 9.51         & 10.01        & 18.41            & 17.04        & 24.65        \\[-.5ex]
               & \BARqhgGRAY{0.8} & \BARqhg{1.0} & \BARqhg{1.1} & \BARqhgGRAY{1.1} & \BARqhg{1.0} & \BARqhg{1.4} \\
          1e-4 & 3.89             & 4.82         & 5.01         & 10.31            & 9.76         & 12.99        \\[-.5ex]
               & \BARqhgGRAY{0.8} & \BARqhg{1.0} & \BARqhg{1.0} & \BARqhgGRAY{1.1} & \BARqhg{1.0} & \BARqhg{1.3} \\[1ex]
               &
          \multicolumn{3}{c}{\TABLETITLE CESM}
               &
          \multicolumn{3}{c}{\TABLETITLE Nyx}
          \\
               &
          \bfseries\textit{qg}
               &
          \bfseries\textit{qh}
               &
          \bfseries\textit{qhg}
               &
          \bfseries\textit{qg}
               &
          \bfseries\textit{qh}
               &
          \bfseries\textit{qhg}
          \\[1ex]
          1e-2 & 61.21            & 24.24        & 75.50        & 118.94           & 30.24        & 164.39       \\[-.5ex]
               & \BARqhgGRAY{2.5} & \BARqhg{1.0} & \BARqhg{3.1} & \BARqhgGRAY{3.9} & \BARqhg{1.0} & \BARqhg{5.4} \\
          1e-3 & 20.78            & 18.38        & 28.13        & 28.25            & 23.92        & 40.17        \\[-.5ex]
               & \BARqhgGRAY{1.1} & \BARqhg{1.0} & \BARqhg{1.5} & \BARqhgGRAY{1.2} & \BARqhg{1.0} & \BARqhg{1.7} \\
          1e-4 & 9.98             & 10.29        & 12.50        & 12.87            & 15.27        & 17.95        \\[-.5ex]
               & \BARqhgGRAY{1.0} & \BARqhg{1.0} & \BARqhg{1.2} & \BARqhgGRAY{0.8} & \BARqhg{1.0} & \BARqhg{1.2} \\
     \end{tabular}%
}
	\caption{Averaged compression ratios (per dataset) of different compression schemes on 109 fields of 4
		datasets with 3 error bounds of $10^{-2}$, $10^{-3}$, $10^{-4}$ (relative to value range). \textit{q} denotes {\quantcode}
		as starting point, \textit{h} denotes customized variable-length encoding (multi-byte-symbol Huffman coding),
		\textit{g} denotes gzip-featured scheme. ``\textit{ab}'' denotes scheme \textit{a} precedes scheme \textit{b}.
	}
	\label{tab:rr-qhg}
\end{table}

\TAB~\ref{tab:rr-qhg} shows the compression ratio variances by applying the {\cusz} workflow followed by \texttt{gzip}.
\emph{q}, \emph{h}, \emph{g} denote \emph{prediction-\textbf{q}uantization}, \emph{multi-byte \textbf{H}uffman coding},
\texttt{\textbf{g}zip}, respectively, and the letter sequence indicates the order of processes (e.g., \emph{qh}
indicates that \emph{h} comes after \emph{q}). However, this additional \texttt{gzip} so far does not exist in {\cusz}
but can demonstrate the potentially achievable compression ratio by exploiting the repeated symbol pattern. More
specifically, by changing the error bound from $10^{-4}$ to $10^{-2}$, Lorenzo predictor generates the intermediate
	{\quantcode}s that exhibit stronger repeated patterns, as indicated in \emph{qh}. For example, when changing from
\emph{qh} to \emph{qhg}, HACC data shows only a 1.04$\times$ improvement in compression ratio under the error bound of
$10^{-4}$, while the compression ratio is improved by 1.52$\times$ under the error bound of $10^{-2}$. We use the
compression ratio of this \emph{qhg} as a reference in the following discussion.

\subsubsection{Limitation of {\cusz}'s Decompression Performance}
{\cusz} follows CPU-SZ's scheme to sequentially reconstruct the prediction values in decompression (per data chunk).
Specifically, the reconstructed value of one item must rely on its preceding values that are fully reconstructed.
As a result, this scheme naturally has a sequential implementation because of data dependency.
Moreover, since both CPU-SZ and {\cusz} store the unpredicted data and the prequantized data separately, it introduces an extra handling step involving if-branch in decompression, which impedes fine-grained data parallelization.
In addition, compared to \cusz's compression kernel, its decompression kernel's throughput is relatively low \cite{cusz2020}.
In particular, the Lorenzo construction kernel can achieve the same order of magnitude of throughput as memory copy \cite{cusz2020}, while the Lorenzo reconstruction kernel has one order of magnitude less in throughput.
All the above prevent {\cusz} from broader use scenarios such as in-situ compression.

\subsubsection{Importance of Lorenzo Predictor}
The modular design of SZ enables adaptively adopting various predictors for different scientific uses.
Among all predictors, the first-order Lorenzo predictor plays an essential role in the SZ framework and is the default predictor since it achieves relatively low prediction error in most cases, as proven in prior works \cite{sz17,liang2018error,zhao2021optimizing}.

Overall, in this work, we endeavor to significantly boost the compression ratio and (de)compression performance of {\cusz}
(e.g., Lorenzo reconstruction kernel) by developing a series of optimization techniques to address the above issues.

\begin{figure*}[!htbp]
	\small\centering
	\includegraphics[width=\linewidth]{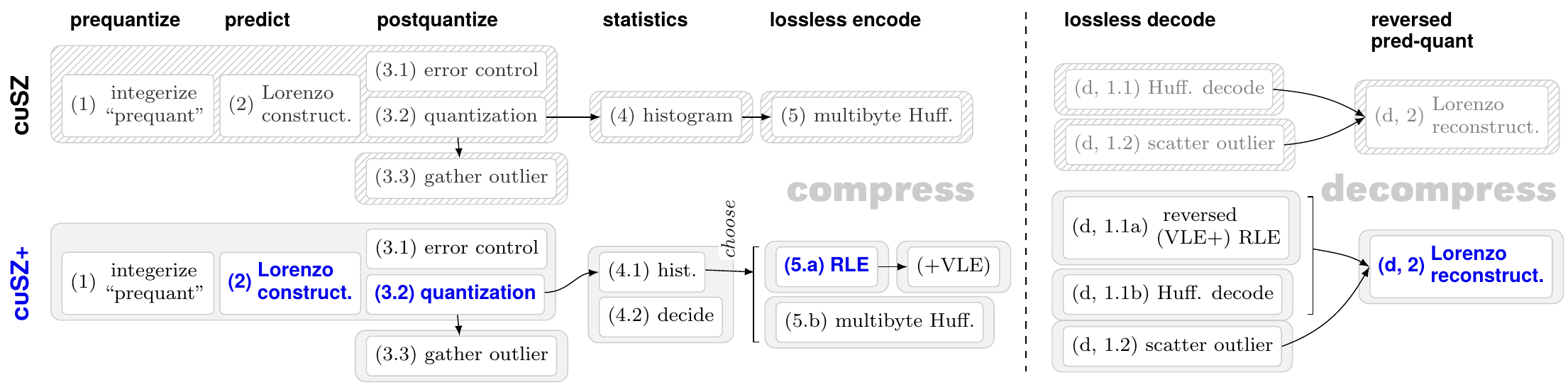}
	\caption{Compression (left) and decompression (right) workflows of the original {\cusz} (top, line patterned) and
		our {\cuszx} (bottom). We design an adaptive solution toward better throughput and compression ratio, featuring 2
		workflow paths. The white block indicates functionality; the parenthesized number marks executing order, where the
		additional letters ``a'' and ``b'' mark the two paths to choose from; the gray enclosure indicates GPU kernel; the arrow
		indicates memory copy. The changes from {\cusz} to {\cuszx} are marked with blue boldface.}
	\vspace{-4mm}
	\label{fig:overview}
\end{figure*}

\section{Compressibility-Aware Framework on GPU}
\label{sec:c13y-aware}

In this paper, we propose a compressibility-aware framework that can significantly improve compression ratios.
In {\cusz} \cite{cusz2020}, all the computations are executed on GPU for high-performance purposes, leading to compression ratios no greater than 32 (for single-precision, or 64 for double-precision).
Such an upper bound is due to the lack of dictionary coding or other pattern-finding-based coding methods.
We note that if ``\emph{data being smooth enough}'' is satisfied, we can apply the alternative run-length encoding (RLE) technique to achieve a higher compression ratio while maintaining 1) the same data quality and 2) a comparable throughput.
In the following text, we use \emph{Workflow-Huffman} to denote the default ``Lorenzo \& multi-byte VLE'' and \emph{Workflow-RLE} to denote ``Lorenzo \& RLE with optional VLE''.

In the following discussion, we first overview our comp\-ressi\-bility-aware design and then give details of our
optimization strategies.
\FIG~\ref{fig:overview} presents an overview comparison between our novel compression framework,
{\cuszx} and the previous {\cusz}. The adaptivity of {\cuszx} is reflected in two workflow paths.

\subsection{Compressibility}\label{sub:c13y}

\subsubsection{Source of Compressibility}

The rationale for the SZ framework to achieve high compression is twofold. First, integer data are easier to compress than IEEE-754 floating-point data.
A non-special \texttt{float}-type number (e.g., non-zero) requires full 32 bits to represent (\texttt{double} requires 64 bits).
SZ's prediction-quantization step transforms prediction errors into {\quantcode}s in integer and eliminates the randomness in terms of floating-point mantissa.
Second, the quant-codes within a predefined range would be compressed further in a lossless manner (i.e., Huffman encoding in our case), while the out-of-range prediction errors are \emph{outliers}.
Then, we enumerate the in-range {\quantcode}s as bin numbers of the histogram and later as symbols in Huffman codebook.
Without context, generic lossless compressions interpret the input as a stream of bytes.
In contrast, the enumeration in a power-of-two $n_\text{enum}$ can exceed 256 and thus
overflow a byte. So, we use at least $\lceil n_\text{enum}/{8} \rceil$ bytes to represent symbols, which can be
single-byte or multi-byte.
The byte interpretation ensures that the enumeration reflects the distribution of {\quantcode}s such that in
Huffman coding, frequent symbols are encoded with fewer bits.

\subsubsection{Reference Compression Ratio}
\label{subs:ref-cr}

\TAB~\ref{tab:rr-qhg} enumerates the \textit{possible} lossless compression techniques that are from CPU-SZ and \cusz
and show consequential compression ratio with different error bounds quantitatively. In the table, \textit{qg} serves to
demonstrate a \textit{presumed} suboptimal scenario; the single-byte interpretation (by a generic lossless compressor) does not indicate the most likely {\quantcode} and hence hurt the compressibility.
\textit{h} indicates the Huffman coding of multi-byte symbols used in both CPU-SZ and \cusz.
\textit{qh} indicates {\cusz} compression schemes that are done on GPU entirely, while \textit{qhg} adds \textit{gzip} to exhibit the highest possible compression ratio, which is archived by CPU-SZ.
We will use the compression ratio from \textit{qhg} as a reference in the following discussion.

\subsubsection{Data Feature Awareness}

Even though it is possible to achieve optimal compression ratio by appending another pattern-exploiting stage to \cusz, it affects the throughput severely since \texttt{gzip} takes place on \emph{host}.
This motivates us to visit the data features that can infer compressibility.
On the other hand, utilizing the repeated pattern is non-trivial because pattern-finding is usually implemented in dictionary coding with irregular accesses and has been reported as low in throughput:
for instance, LZ4 features relatively high throughput but still can be significant in latency as one
appended stage \cite{nvcomp-lz4}.

In this work, we propose a solution to exploit the repeated data pattern by leveraging the indication of data
smoothness.
The prediction would generate two types of data: zeros and non-zeros.
The zero represents the cases in which the prediction error is no greater than one unit of the error bound ($eb$) regarding the original value. And the non-zeros represent the otherwise, which are expected scarce and scattered across the data.
With such dichotomy, we consider that the {\quantcode}s are smooth when they are \emph{locally} continuous with zeros and propose to use run-length encoding to utilize the data patterns that may exist.

Optionally, we can append another stage of Huffman encoding, which can further bring a steady 2$\times$ to 3$\times$
ratio gain beyond RLE.
Our design goal is to push the compression ratio beyond the original
32$\times$ for \texttt{float} (or 64$\times$ for \texttt{double}); considering the throughput, compressing the metadata
of RLE output is optional and by default disabled in GPU processing.
We further expand the criteria and the uses of RLE in \SEC\ref{subs:rle}.

\begin{figure}[ht]
	\centering
	\begin{subfigure}[b]{\linewidth}
		\centering
		\includegraphics[width=.8\linewidth]{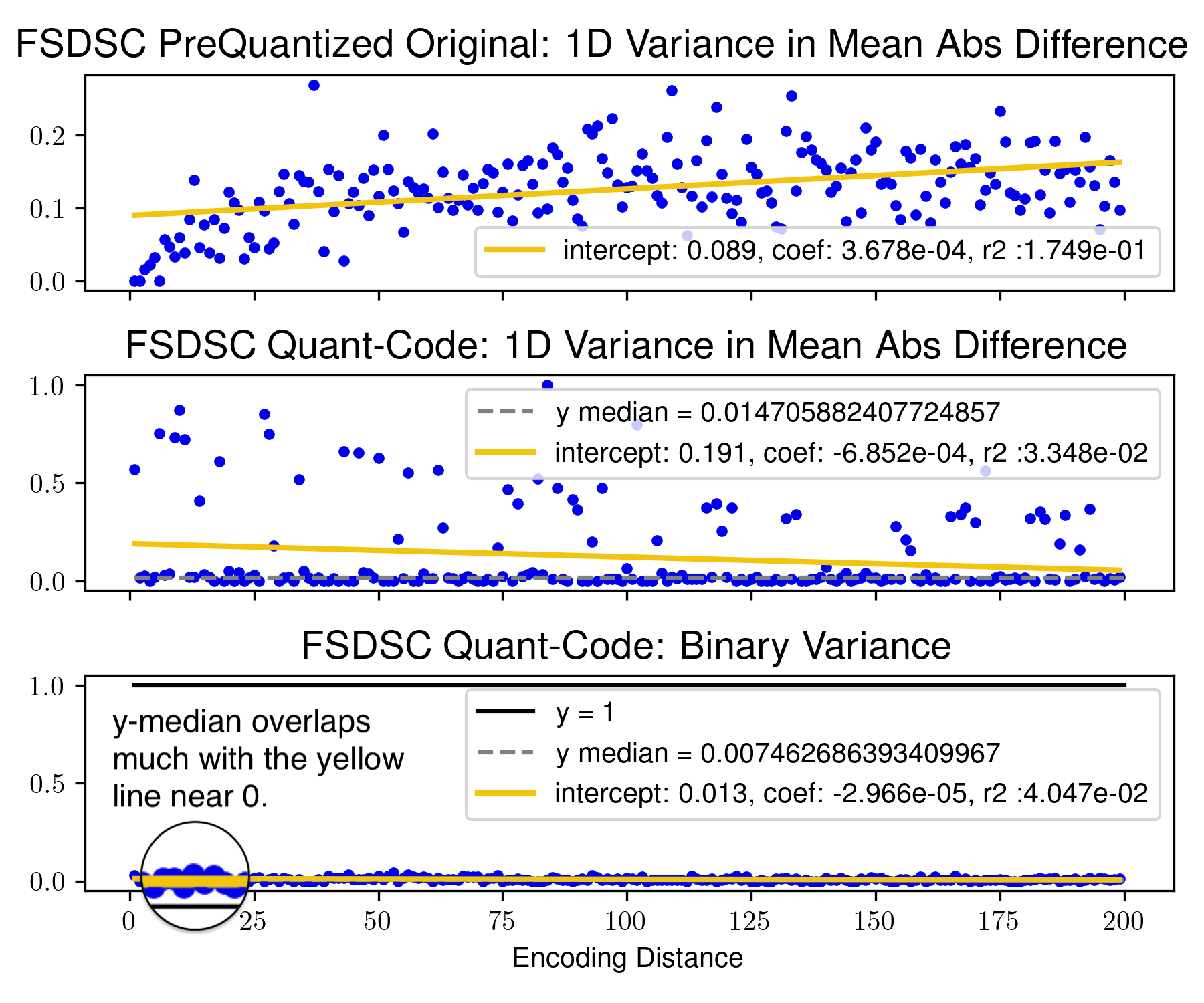}
		\caption{Smoothness against encoding distance (\texttt{CESM FSDSC} at 1e-2). The yellow line denotes linear
			regression of variances at distances.}
		\label{subfig:variance}
	\end{subfigure}
	\begin{subfigure}[b]{\linewidth}
		\centering
		\includegraphics[width=.8\linewidth]{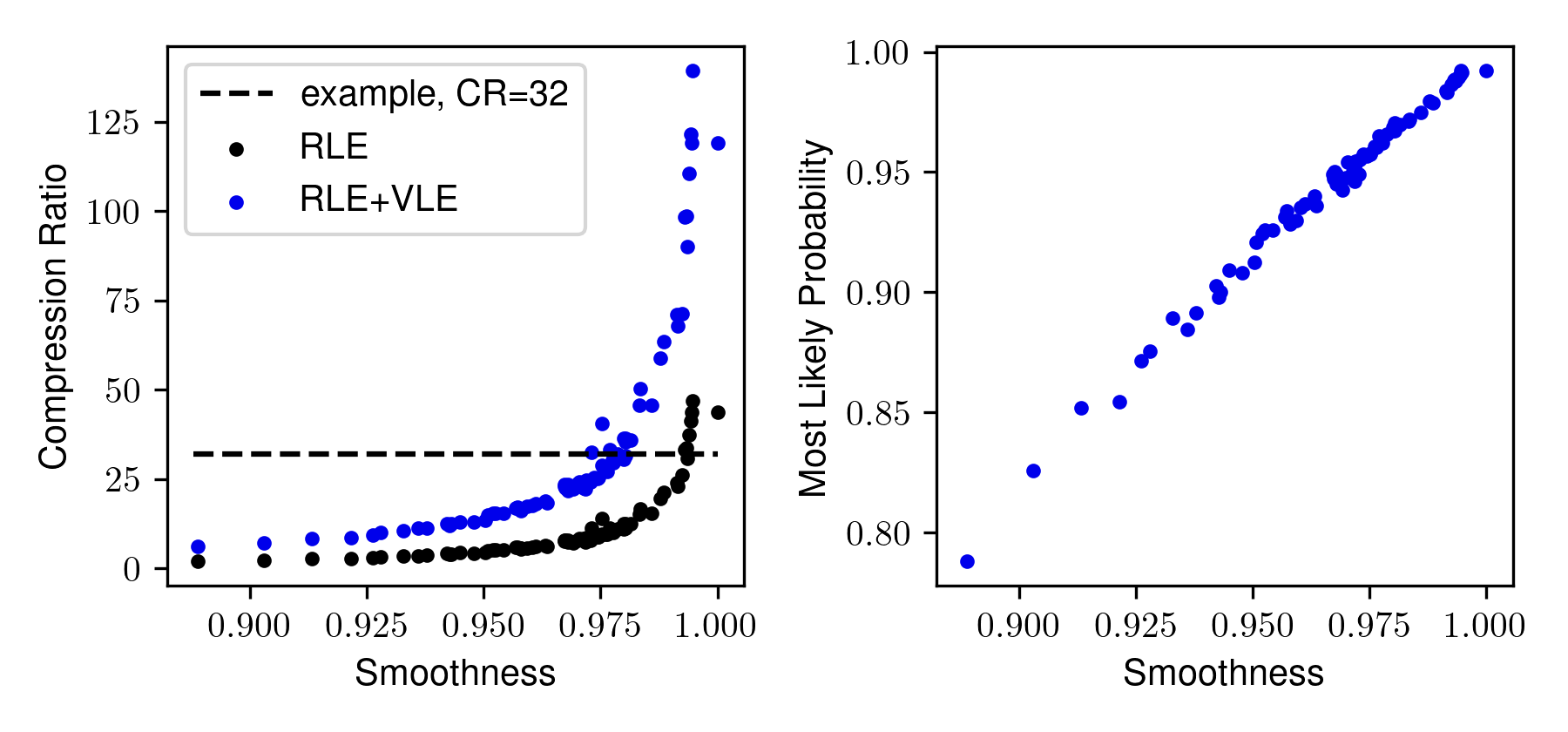}
		\caption{Smoothness-Probability of the most likely symbol relationship.}
		\label{subfig:relationship}
	\end{subfigure}
	\caption{Smoothness of prequantized data and {\quantcode}, and smoothness-probability of the most likely symbol relationship can help determine when to use RLE. For example, a threshold compression ratio can be set to 32 to find the desired smoothness, and the smoothness can be directed to the probability of the most likely symbol.}
	\label{fig:madogram}
	\vspace{-4mm}
\end{figure}

\subsection{Smoothness and Run-Length Encoding}
\label{subs:rle}

RLE, first introduced in \cite{rle1967}, is a form of lossless compression, in which sequences of consecutive same-value data elements are stored as value-count tuples. Such kinds of sequences are called \textit{runs} of data.
For example, ``aabcccccaa'' is stored as ``(a,2)(b,1)(c,5)(a,2)''. RLE's continuity in the same values can be seen as a simplistic pattern-finding method; its regular access when checking the following values can contribute to the high throughput on GPU.
Without the prefix property of Huffman coding, the length of runs must be recorded as metadata, introducing overhead.
The overhead immediately raises questions of (1) how to model the compressibility and (2) when to use RLE.
We identify that madogram, a variogram \cite{cressie1980robust} variant, and histogram can help make the decision---madogram suffices to reveal the reason RLE can perform well, and histograming is easy to conduct and coincide can converge to indicate the chances of performing RLE.

\subsubsection{Estimation from Histogram}
We first discuss the data features that motivate VLE, based on the histogram.
The entropy value of the histogram is calculated as
$H(X) = -\sum p_i\log_2 {p_i}$,
where $p_i$ is the probability of the $i$-th symbol.
We use $\avgbits$ to denote the average bit-length of Huffman codeword, and $\langle b \rangle_\text{RLE}$ to denote that of RLE. We use $R$ (redundancy) to denote the discrepancy
between $\avgbits$ and entropy $H(X)$ (i.e., $R = \avgbits - H(X)$).
Denote the probability of the most likely symbol by $p_1$.
With $p_1$, we can estimate the upper and lower bounds of $R$, $R^+$ and $R^-$, respectively.
When $p_1 > 0.4$, $R^-$ is given by $1 - H \left(p_1, 1-p_1\right)$, where $H\left(p_1, 1-p_1\right) = \textstyle p_1\log_2 \frac{1}{p_1} + \left(1-p_1\right)\log_2 \frac{1}{1-p_1}$ \cite{huffman-redundancy}.
The upper bound is given by $R^+ = p_1+0.086$ (no restriction) \cite{huffman0086}.
Therefore, without building Huffman tree, we can estimate the upper and lower bound of $\avgbits$ with $R^+$ and $R^-$, respectively.
A lower bit-length leads to a higher compression ratio.
We expect to use RLE when $\langle b \rangle_\text{RLE} \le \avgbits$. And we also use the upper bound of $\avgbits$ to estimate the lowest gain of an additional Huffman coding after RLE (see \TAB~\ref{tab:rle-better-than-vle}).

\subsubsection{Modeling RLE Compressibility}

Due to the obvious overhead from storing the metadata, RLE would do when the penalty caused by value change is sufficiently low. And the low penalty can be translated to the data being \emph{smooth enough}.
However, the histogram cannot reflect the smoothness because a locally smooth datum can have a similar histogram to a rougher one.

The method of variogram \cite{cressie1980robust} inspires us to derive a new scheme to measure the smoothness. Variogram
(i.e., general-purpose multidimensional data variance) is a very effective metrics to reveal a
variance-distance relationship in spatial data based on sampling. Its theoretical form is
\[
	\begin{split}
		2\gamma \left(\mathbf {s} _{1},\mathbf {s} _{2} \right)
		= {\operatorname{var}} \big( Z(\mathbf {s} _{1})-Z(\mathbf {s} _{2}) \big)
		= E\left[ \big( Z(\mathbf {s} _{1})-Z(\mathbf {s} _{2}) \big)^{2}\right],
	\end{split}
\]
where $Z(\mathbf{s})$ is a spatial random field. Considering the encoding iteration is unidimensional, we substitute the
power description $\big(Z(\mathbf{s}_{1}) - Z(\mathbf{s}_{2})\big)^{2}$ with the absolute difference $| Z(\mathbf {s}
	_{1})-Z(\mathbf {s} _{2})|$ to form \emph{madogram}. Also note that an \emph{RLE run} discontinues when the
value differs from the current one ($v_\text{this}$), we further adjust the absolute difference to \emph{binary}
difference, defined as
\[\text{binary variance} =
	\left\{\begin{array}{cc}
		0 & v_\text{this} = v_\text{next}    \\
		1 & v_\text{this} \neq v_\text{next}
	\end{array},
	\right.
\]
regardless of the distance.
And the expected value (defined below) is interpreted as \emph{RLE roughness}; then, \emph{smoothness} is naturally $(1-\text{roughness})$.
Given the $\mathcal{O}\left(n^2\right)$ nature of enumerating pairwise variances, the empirical madogram method would do with an offline sampling scheme.
More specifically, given a sufficiently large number sampling number $N$ and a maximum distance of measurement $D_\text{max} = 200$, we form the pair $(a, a+d)$, where $a$ is randomly selected from the whole data field, and $d = \operatorname{rand}(1, 200)$ (suppose $(a+d)$ is in the data range).
The summed variance of each distance is averaged by its corresponding count.
The averaged binary variance $v(d)$ remarks the roughness, and $1-v(d)$ the smoothness.

We first show the madogram of the prequantized original data and {\quantcode} in absolute difference and the madogram of {\quantcode} using binary variance in \FIG\ref{fig:madogram}.
\FIG\ref{subfig:variance} shows that {\quantcode} is smoother with less variance than the prequantized original data.
That is to say, the prediction-quantization scheme uses much less information to represent the data change and therefore helps achieve a high rate of data reduction.
The third part of \FIG\ref{subfig:variance} indicates that {\quantcode} can forward-encode from a fixed starting point with almost equal roughness at an arbitrary distance from the starting point.
Hence, at a stable rate of roughness, it is worth performing RLE.
The next step is to determine the threshold.
\FIG\ref{subfig:relationship} shows that with binary variance we can relate (1) data continuity/smoothness and the compression ratio (CR) and (2) data smoothness and the probability
of the most likely symbol ($p_1$). With (1), for example, a simplistic case is to set a CR threshold at 32 and look up
the smoothness for RLE or RLE+VLE. With (2), $p_1$ can determine compression ratios of both the proposed RLE workflow
and VLE in {\cusz} and be used to select from the two workflows. For example, \texttt{FSDSC} has an RLE-CR above 25
while {\cusz}-VLE-CR is 26$\times$ to 29$\times$. Note that 1) there is an overhead of chunkwise metadata from {\cusz}-VLE (the
final CR is 23.88) and 2) additional VLE after RLE can provide steady 3$\times$ more CR in general (estimated from the
corresponding histogram), making the accumulated CR above 70$\times$ in this case.

We can use the mapping to determine when to use RLE. For example, we can set a threshold of 32$\times$, the highest possible compression ratio obtained from Huffman coding, and find the empirical smoothness hence the proper $p_1$.
Or, more conveniently, we can give a practical conclusion: when Huffman is likely to achieve an average bit-length lower than 1.09, we can use RLE.

\section{Performance Optimization}
\label{sec:design}

In this section, we present our optimization strategies, featuring (1) performance improvements on \emph{each} kernel in
compression, and (2) a new high-performance Lorenzo reconstruction kernel in decompression.

\subsection{Compression Optimization}
\label{sub:const-sparsity-aware}

\subsubsection{Dual-Quantization}

In Original SZ, in situ data reconstruction is required during compression; such reconstruction is precisely the same as decompression-time one.
Specifically, a data item $d$ is
reconstructed from the known previous items, based on their known predecessors recursively. Such process 1) makes a
compression-time reconstructed item in place of the original, iteratively, and therefore 2) causes loop-carried
\emph{read-after-write} dependency. And {\quantcode} that controls the error compensation is one outcome of this
process, which would be encoded further. To get {\quantcode} $q$ and use it to reconstruct data $d$ in an arbitrary
iteration, SZ needs to go through the following data transformation. I) the prediction error is from $e^\circ=d -
	p^\circ$, where $p^\circ$ denotes predicted value. II) with respect to the user-input error bound $eb$, $e^\circ$ is integerized to {\quantcode} $q^\circ$ with rounding. III) $e^{\circ\star}$ is transformed from $q^\circ$ and serves as
\emph{error compensation} to $p^\circ$ such that $d^{\circ\star} = e^{\circ\star} + p^\circ$ approximates $d$ with a
loss that is no greater than $1\times eb$, ensuring error-boundness.

\paragraph{Generality}
{\cusz} work \cite{cusz2020} resolved the tight RAW dependency by foregoing integerization. Its essential technique is
two-phase \textit{dual-quant}, including
\begin{description}
	\item[\sffamily\bfseries prequant] With integerization $d^\circ = \operatorname{round}\langle d/(2\cdot eb)\rangle$,
		where $d$ is the original, the compression error $
			|d-d^\circ\cdot 2eb| < eb$ is guaranteed.
	\item[\sffamily\bfseries postquant]
		The difference between the prediction $p^\circ$
		and the target integer value $d^\circ$ is rendered as $\delta^\circ = d^\circ - p^\circ$.
		The {\quantcode} $q^\circ$ is equivalent to but typecasted from $\delta^\circ$.
\end{description}
Note that $\delta$ is the counterpart of error compensation $e$; the mark $\delta$ is deliberately chosen for there is no
further error introduced after prequant. Unlike error compensation $e^\circ\neq e^{\circ\star}$ in Original SZ, $d^\circ
	{\to} \delta^\circ$ and $\delta\equiv q$ guarantees the reconstructed $d^{\circ\star} = p^\star + \delta = d^\circ$.
Therefore, it obviates the calculation after $q^\circ$. And $d^\circ$, equivalently the known reconstructed data, is
ready after prequantization, eliminating the loop-carried RAW dependency parallelizing prediction-quantization.

\paragraph{Computational Efficiency}
Computation-wise, it is worth noting that {\cusz}'s dual-quant method continues working for {\cuszx} with Lorenzo
predictor and a modified quantization scheme. According to Tao~et~al. \cite{sz17}, the general-form Lorenzo predictor is
given by\\
$\textstyle
	\sum
	^{{k_{1\ldots d}\neq\mathbf{0}}}
	_{{0\le k_{1\ldots m} \le n}}
	\left\langle\prod^m_{j=1}(-1)^{k_j+1} {n \choose k_j} \right\rangle \cdot d_{x_1 - k_1,\,\cdots,\,x_d - k_d}$, where \\
$\textstyle
	\sum^{{k_{1\ldots d}\neq\mathbf{0}}}_{{0\le k_{1\ldots m} \le n}}
	\left\langle\prod^m_{j=1}(-1)^{k_j+1} {n \choose k_j} \right\rangle = 1$,
that is, through\-out the prediction, coefficients sum to 1. Thanks to \textit{dual-quant}, the integer coefficient is a set
$C=\{c \mid c\in \mathbb{N}\}$ that is closed under addition, subtraction, and multiplication (i.e., no division
involved). Moreover, integer-based data reconstruction is precise and robust with respect to machine $\epsilon$. In
addition, integer summation is considered as commutative, i.e., $a\oplus b=b\oplus a$ and  $\oplus$ becomes
integer addition. Thus, given arbitrary numbers of integers, adjusting the addend order results in no difference in sum.
This property guarantees that our proposed fine-grained Lorenzo reconstruction (will be discussed in \SEC\ref{sub:tpds-reconstruction}) can reorder the prediction computation.

\subsubsection{Compression Kernel Enhancement}

We mainly focus on optimizing two kernels: Lorenzo construction and the Huffman encoding.
Besides coalescing interaction with DRAM/shared memory, we propose to adopt two primary strategies to increase the number of
thread blocks (or warps) that can run concurrently within one SM (streaming multiprocessor). I) We coarsen the
granularity by assigning more data items to one thread. For example, a 16$\times$16 2D data chunk is equally split
into two groups, each traversed in consecutive 8 items along $y$-direction. Note that the data-thread
mapping my differ from the coalescing load~\& store. Then, it is possible to launch more warps per SM toward higher occupancy.
II)
According to the extrapolative prediction form, neighboring data items are reused, with the index difference being 1.
We perform in-warp shuffle to exchange data.
This strategy can decrease the shared memory use to launch more warps in the
same SM. The comparison between the kernels of {\cusz} and our {\cuszx} is shown in \TAB~\ref{tab:v100-cmp}.

\subsection{Decompression Optimization}
\label{sub:tpds-reconstruction}

\subsubsection{The Modified Quantization Scheme}

We first modify the quantization scheme of compression to eliminate the divergence in the reconstruction procedure, enabling fine-grained parallelism on GPU architectures.
In {\cusz}, if the error compensation
$\delta^\circ$ at the {prequant}ized $d^\circ$ is out-of-range, $d^\circ$ is otherwise stored as an
\texttt{outlier}, with 0 stored as the placeholder. In {\cuszx}, if the compensation $\delta^\circ$ is
out-of-range, $\delta^\circ$ is stored as an \texttt{outlier} (line 7 in Algorithm~\ref{algo:const-reconst}), while the
	{\quantcode} remains stored in the same way (line 5 in Algorithm~\ref{algo:const-reconst}).
Then, the outlier and the
	{\quantcode} are further processed, i.e., stored directly and compressed in a lossless manner, respectively. During the
decompression, the outlier and the {\quantcode} are extracted from the compression archive as it is and
from lossless decoding, respectively.

During decompression, the original {\cusz} accesses \texttt{outlier} when hitting 0 (the placeholder). However,
{\cusz}'s coarse-grained reconstruction only exploits the parallelism of multithreading but rarely considers dependency, memory
access pattern, and computational efficiency. In comparison, by modifying quantization, {\cuszx} fuses the
	{\quantcode} and the outlier before reconstruction (line 12 in Algorithm~\ref{algo:const-reconst}). In such a
manner, we conduct the reconstruction from the error compensation $\delta^\bullet$ without any stall, hence eliminating
the dependency that exists in {\cusz}'s coarse-grained reconstruction.

\begin{algorithm}[t]
	\setstretch{1.15}
	\centering
	\caption{Lorenzo construction and reconstruction. Yellow-highlight marks the modified quantization scheme.
		Blue-highlight marks partial-sum based reconstruction.}
	\label{algo:const-reconst}
	\footnotesize\ttfamily\setstretch{1.5}
\begin{algorithmic}[1] 
    \State (for all fp-prepresented data item $d$)
    \Comment{{\color{Bv}\fontfamily{ugq}\selectfont compression}}
    \State \makebox[1.25em][l]{$d^\circ$}%
    \makebox[6em][l]{$\gets (d)$.divided\_by($2\!\times\!eb$)}%
    \Comment{\textbf{pre}quant, barrier}
    \State \makebox[1.25em][l]{$p^\circ$}$\gets\ell(d^\circ_\text{SR})$,\quad$\delta^\circ \gets p^\circ\! - d^\circ$
    \If{$\delta^\circ <$ cap/2 $\equiv$ radius $r$}  \Comment{\textbf{post}quant}
    \State \hspace*{-\fboxsep}\colorbox{Y!40}{%
        $q^\circ \gets$ $(\delta^\circ).\text{to\_int()}+r$%
    }
    \Comment{captured, to lossless-compress}
    \Else
    \State \hspace*{-\fboxsep}\colorbox{Y!40}{outlier $\gets \delta^\circ$}
    \Comment{remaining fp presence}
    \EndIf
    \Statex
    $p\Sigma$ to denote inclusive partial-sum
    \Comment{{\color{Bv}\fontfamily{ugq}\selectfont decompression}}
    \Statex (for all $q^\bullet\equiv q^\circ$)
    \State \hspace*{-\fboxsep}\colorbox{Y!40}{$q^{\mathclap{\,\,\prime}\phantom{\bullet}} \gets (q^\bullet \oplus \text{outlier}) - r$}
    \Comment{fuse quant. and outlier}
    \State
    \hspace*{-\fboxsep}\colorbox{Bv!20}{$d^\bullet\gets p\Sigma_x\,q^\prime$ only if dim. $x$ exists}
    \Comment{barrier}
    \State
    \hspace*{-\fboxsep}\colorbox{Bv!20}{$d^\bullet\gets p\Sigma_y\,(p\Sigma_x\,q^\prime)$ only if dim. $x$, $y$ exist}
    \Comment{barrier}
    \State
    \hspace*{-\fboxsep}\colorbox{Bv!20}{$d^\bullet\gets p\Sigma_z\,(p\Sigma_y\,(p\Sigma_x\,q^\prime))$ only if dim. $x$, $y$, $z$ exist}
    \Comment{barrier}
    \State output $\gets d^\bullet\cdot (2\!\times\!eb)$
\end{algorithmic}
	\vspace{-5mm}
\end{algorithm}

\subsubsection{Partial-Sum Lorenzo Reconstruction}
\begin{figure*}[ht]
	\begin{subfigure}[t]{0.39\textwidth}
		\centering
		\includegraphics[width=1\linewidth]{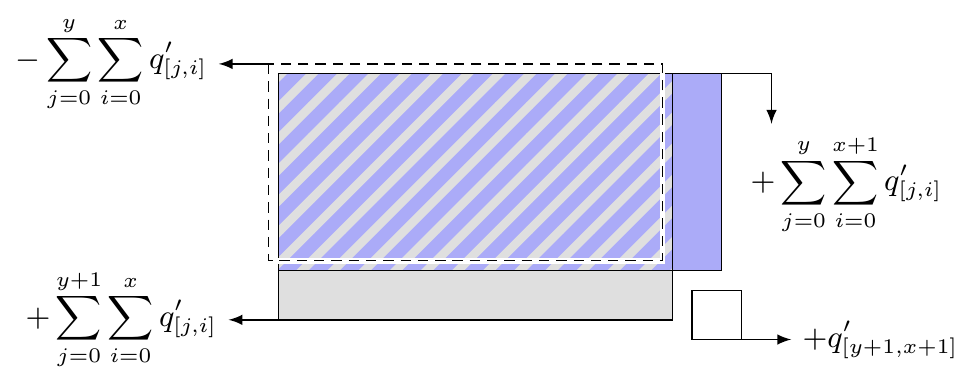}
		\caption{Concept of reconstruction in terms of 2D partial-sum.}
		\label{fig:2d-partial-sum}
	\end{subfigure}%
	~%
	\begin{subfigure}[t]{0.6\textwidth}
		\centering
		\resizebox{1\linewidth}{!}{\begin{tikzpicture}

    \node{\small\color{gray}
        \begin{tabular}{@{} cc @{}}
            \color{black} partial-sum along $x$ & \color{black} partial-sum along $y$                                                                                                        \\[1.5ex]
            $\PartialSum{x\!-\!1}{y-2}           \xrightarrow{\QuantAddend{y-2}{x-1}} \PartialSum{x}{y-2}$
                                                & $\PartialSum{\PartialSum{x}{y-2}}{x}           \xrightarrow{+\PartialSum{x}{y-2}}           \PartialSum{\PartialSum{x}{y-1}}{x}$           \\[2ex]
            $\PartialSum{x\!-\!1}{y-1}           \xrightarrow{\QuantAddend{y-1}{x-1}} \PartialSum{x}{y-1}$
                                                & $\PartialSum{\PartialSum{x}{y-1}}{x}           \xrightarrow{+\PartialSum{x}{y-1}}           \PartialSum{\PartialSum{x}{y\phantom{-0}}}{x}$ \\[2ex]
            $\PartialSum{x\!-\!1}{y\phantom{-0}} \xrightarrow{\QuantAddend{y\phantom{-0}}{x-1}} \PartialSum{x}{y\phantom{-0}}$
                                                & $\PartialSum{\PartialSum{x}{y\phantom{-0}}}{x} \xrightarrow{+\PartialSum{x}{y\phantom{-0}}} \PartialSum{\PartialSum{x}{y+1}}{x}$           \\[2ex]
        \end{tabular}
    };


\end{tikzpicture}}
		\caption{Exemplary 2-pass partial-sum computation for 2D data reconstruction.}
		\label{fig:2-time}
	\end{subfigure}

	\label{fug:nd-partialsum}
	\caption{Example of 2D partial-sum computation for Lorenzo reconstruction in \cuszx's decompression.}
\end{figure*}
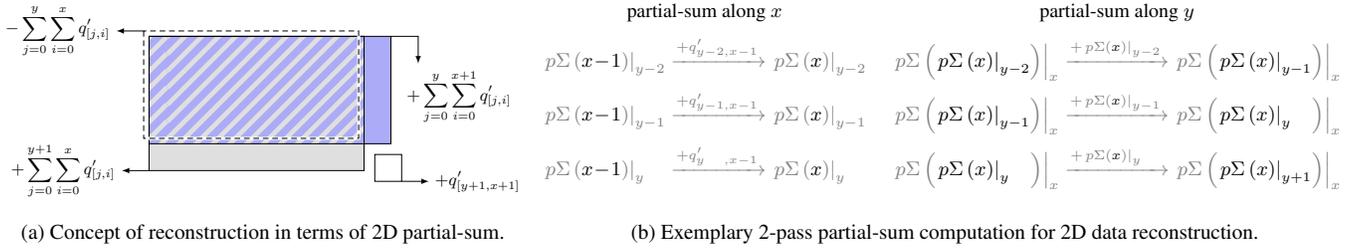

The default 1D to 3D first-order Lorenzo predictors are put as follows,
{\footnotesize
		\[
			\begin{split}
				p_\TEXT{x} =& + d_\TEXT{x-1}                                          \\
				p_\TEXT{y,x} =& - d_\TEXT{y-1,x-1} + d_\TEXT{y-1,x} + d_\TEXT{y,x-1}  \\
				p_\TEXT{z,y,x} =& + d_\TEXT{z-1,y-1,x-1} - d_\TEXT{z-1,y-1,x}
				- d_\TEXT{z,y-1,x-1} + d_\TEXT{z,y-1,x}                               \\
				& - d_\TEXT{z-1,y\phantom{-1} ,x-1} + d_\TEXT{z-1,y\phantom{-1} ,x}
				+ d_\TEXT{z,y\phantom{-1} ,x-1}
			\end{split}
		\]
	}%
In the following text, we use 2D form to demonstrate the expression. Let $r$ denote quantization radius. In
decompression, $d^\bullet = p^\bullet + q^\bullet\! - r$; with $q^\prime = q^\bullet\! - r$, it becomes $d^\bullet =
	p^\bullet + q^\prime$. Considering that we initially predict from zeros, the first predicted item is
$d^\bullet_\TEXT{0,0} = q^\prime_\TEXT{0,0}$. We then observe that an arbitrary item $\TEXT{y,x}$ is predicted as
$\sum^\TT{y}_{\TT{j=0}}\sum^\TT{x}_{\TT{i=0}}q^\prime_\TEXT{j,i}$. We give a proof by induction on $\TT{[y,x]}$, as
$d^\bullet_\TEXT{y+1,x+1}$ equals to
	{\footnotesize
		\[
			\begin{split}
				& - \sum^\TT{y}_{\TT{j=0}}\sum^\TT{x}_{\TT{i=0}}q^\prime_\TEXT{j,i}
				+ \sum^\TT{y}_{\TT{j=0}}\sum^\TT{x+1}_{\TT{i=0}}q^\prime_\TEXT{j,i}
				+ \sum^\TT{y+1}_{\TT{j=0}}\sum^\TT{x}_{\TT{i=0}}q^\prime_\TEXT{j,i}
				+ q^\prime_\TEXT{y+1,x+1}   \\
				=&\phantom{+\ } \sum^\TT{y}_{\TT{i=0}}q^\prime_\TEXT{j,x+1} + q^\prime_\TEXT{y+1,x+1}
				+ \sum^\TT{y+1}_{\TT{j=0}}\sum^\TT{x}_{\TT{i=0}}q^\prime_\TEXT{j,i}
				= \sum^\TT{y+1}_{\TT{j=0}}\sum^\TT{x+1}_{\TT{i=0}}q^\prime_\TEXT{j,i}.
			\end{split}
		\]
	}%
An intuitive demonstration for 2D case is shown in \FIG\ref{fig:2d-partial-sum}, with canceling the joint summation.
Similarly, the computation for $N$-D case can be done by $N$-D partial-sum as follows.

\paragraph{Computation}

We define $N$-D partial-sum of $x$ till index $[k_N,\ldots,\allowbreak k_2,k_1] \in \mathbb{N}^N$ as
{\small
\[
	p\Sigma(x; k_N,\ldots,k_2,k_1) = \sum^{k_N}_{i_{N}=0} \cdots \sum^{k_{2}}_{i_{2}=0} \sum^{k_{1}}_{i_{1}=0}
	x_{[i_N,\ldots,i_2,i_1]},
\]
}%
where $p\Sigma$ is a variadic operator for any $N$. We can decompose it to $N$-pass 1-D
partial-sums, as
\[
	\begin{split}
		& p\Sigma(x; k_N,\ldots,k_2,k_1) = p\Sigma ( p\Sigma \big(x; k_{N-1},\ldots,k_2,k_1 \big) ; k_{N} ) \\
		=&
		p\Sigma (
		p\Sigma \big(
			\cdots
			p\Sigma \Big(
			p\Sigma \Big( x; k_1 \Big);
			k_2
			\Big) \cdots ; k_{N-1}
			\big); k_N ).
	\end{split}
\]
That is, the output of a partial-sum on $x_m$-direction is the input of that on $x_{(m+1)}$-direction. Given the problem
size $(X_N,\ldots,X_2,X_1)$, where $k_{(\cdot)} \le X_{(\cdot)}$, a pass along $x_{(\cdot)}$ features the degree of
independence (hence the maximum possible parallelism) equal to $\prod_{i\neq (\cdot)} X_i$.

We give an example for 2D case of size $b_y$-by-$b_x$.
The first partial-sum along $x$ is performed through indices
$[y,0\ldots b_x]$, given any $y$; the partial-sum at $\TEXT{y,x}$ is $p\Sigma \left(q^\prime|_y;x \right) =
	\sum^\TT{x}_\TT{i=0} q^\prime_\TEXT{y,i}$. The second partial-sum along $y$ is performed through indices $[0\ldots
			b_y,x]$, given any $x$; the partial-sum at $[y,x]$ is $p\Sigma(q^\prime|_{y,x}; y,x) = p\Sigma \left( \, p\Sigma \left(
		q^\prime|_y;x \right)\!|_x;y\, \right)$. Their parallelism degrees are $b_y$ and $b_x$, respectively. An illustration
of this parallelized computation is given in \FIG~\ref{fig:2-time}.

\subsubsection{Implementation Detail}
\label{subs:poc-implement}

We conduct partial-sum in a chunk-wide manner, as the compression is the same way: no inter-chunk dependency.
To illustrate the effectiveness of this proposed solution, we first have a proof-of-concept
implementation using shared memory and assign 1 item to 1 thread. Compared to the coarse-grained implementation in {\cusz},
this one improves in performance notably (see ``na\"ive'' against ``{\cusz}'' in \TAB~\ref{tab:poc-xlorenzo}).
Also, note that the new
reconstruction kernel has high performance, similar to the fine-grained construction kernel in {\cusz} (see ``{\cusz}''
column).
Our proposed fine-grained solution for {\cuszx} exhibits higher resource utilization than the coarse-grained
parallelization does in {\cusz}.

It is worth noting that the arithmetic intensity of the Lorenzo reconstruction kernel is relatively low with linear
algorithmic time complexity, which tends to be memory-bound. Thus, we consider the two optimizations over
the na\"ive implementation:
\begin{enumerate}
	\item We tune the access pattern to ensure the coalesced load/store from/to global memory.
	\item We increase the \emph{sequentiality} to each thread to balance load/store and computation.
\end{enumerate}
With these optimizations, we form a different thread block management, considering that there is no canonical way to
map the thread to the data throughout the dimensions; instead, the data-thread mapping is more bottom-up.
Among all the abstraction levels from warp (``SIMDness'') to grid, warp convergence is the first concern to address.
On the other hand, our problem features more ``streaming''-style memory access given the linear processing time; there is not much of locality that we can exploit. This is distinct from ``persisting'' style found in, e.g., \texttt{gemm}, whose
algorithmic complexity is much above $\mathcal{O}(n)$. Therefore, the workflow path of \emph{global-memory $\to$ register
	$\to$ shared-memory} is neither efficient nor necessary, given that the register file can hold a certain amount of data and
perform in-warp operations.

\paragraph{1D Implementation} We attribute the chunkwise partial-sum to 1D \texttt{BlockScan} and implement it using
\texttt{NVIDIA::cub} library.
Warp-striped load/store from/to global memory is used to ensure the coalesced read/write.
We test different sequentialities of each thread, with warp-/block-wide scan-and-sync employed.
We use 256 as 1D chunk size in {\cuszx} and the \texttt{vx} field in 1D HACC dataset as an example in \TAB~\ref{tab:poc-xlorenzo}.

\paragraph{2D Implementation} There is no such direct abstraction, however, for higher-dimensional partial-sum---the
multidimensional use of \texttt{cub} is internally linearized to 1D. Thus, we handcraft the 2D reconstruction kernel.
We use $16\times16$ as 2D chunk size in {\cuszx} (the same as {\cusz}). The 1-to-1 thread-data binding and
unconditional use of shared memory (the only explicit cache in GPUs) incurs both low computational utilization of each thread
(and hence warp) and underuse of register file. In-warp operations such as \verb|__shfl_up_sync| allow accessing the
register in the same warp and hence data exchange without using shared memory. We then specify $x$-direction as
warp-shuffling space while setting sequentiality to $y$-direction.
Each thread holds an $n_{(2)}$-length thread-private array \texttt{tp[]}, in which a fragment of partial-sum is
sequentially and trivially done. $\frac{16}{n_{(2)}}$ threads are needed along $y$-direction to complete a
size-16 partial-sum. Of the same $y$, all threads except the last one propagate the last-element value in
\texttt{tp[]} to all the elements in the next thread's private array, sequentially, using shared memory to
exchange.
We identify the sequentiality of 8 results in the optimal throughput under such thread block configuration---a
$(16,2,1)$-block size comprises a warp---at 254.2 GB/s on V100 and 508.6 GB/s on A100 with testing on a sample CESM field.
Note the throughputs are comparable to those from the high-throughput 1D kernel.

\begin{table}
	\centering\footnotesize\ttfamily
\resizebox{\linewidth}{!}{
	\begin{tabular}{@{} >{\bfseries\sffamily\scriptsize}r lr >{\scriptsize\color{Bv}}r @{}r >{\scriptsize\color{Bv}}r @{}r >{\scriptsize\color{Bv}}r @{}}
		\multicolumn{2}{@{} l @{}}{\sffamily throughput}
		            &
		\BOLD\color{Bv} {\cusz}
		            &      &
		\BOLD\color{Bv} ours\ \null
		            &      &
		\BOLD\color{Bv} ours\ \null
		            &
		\BOLD\color{Bv} {A100 adv.}
		\\
		\multicolumn{2}{@{} l @{}}{\sffamily measured in GB/s}
		            &
		\cite{cusz2020}
		            &      &
		\BOLD\color{Bv} (naïve)
		            &      &
		\BOLD\color{Bv} (optim)
		            &
		\BOLD\color{Bv} {over V100}                                                 \\[1ex]
		{1D (HACC)} & A100 & -    &         & 219.8 & +130\% & 504.5 & 1.64$\times$ \\
		            & V100 & 16.8 & +1404\% & 252.6 & +24\%  & 313.1 &              \\[1ex]
		{2D (CESM)} & A100 & -    &         & 182.1 & +179\% & 508.6 & 2.00$\times$ \\
		            & V100 & 58.5 & +239\%  & 198.4 & +28\%  & 254.2 &              \\[1ex]
		{3D (Nyx)}  & A100 & -    &         & 147.9 & +174\% & 405.1 & 1.70$\times$ \\
		            & V100 & 29.7 & +492\%  & 175.9 & +35\%  & 238.1 &              \\
	\end{tabular}%
}
	\caption{Proof-of-concept throughput on V100 for \{1, 2, 3\}-D. The referenced throughput of {\cusz}'s
		Lorenzo reconstruction covers all the fields \cite{cusz2020}. The table shows a single field (i.e., \texttt{vx} in HACC,
		\texttt{CLDHGH} in CESM, \texttt{baryon-density} in Nyx) for demonstration purpose.}
	\vspace{-5mm}
	\label{tab:poc-xlorenzo}
\end{table}

\paragraph{3D Implementation} The 3D problem has an addition to the 2D implementation. Right after the same procedure
in the 2D case, we append an $x$-$z$ transposition of the $x$- and $y$-direction partial-sum result and repeat the previous
$x$-direction partial-sum (with $z$-direction data). Based on trials, we identify that the 8$\times$ sequentiality
results in the best throughput. Yet, the 3D kernel, due to the longer computational process, does not achieve as high
throughput as lower dimensionality. Further evaluation and analysis are listed in \TAB~\ref{tab:eval-default} in
\SEC\ref{sec:eval}.

\section{Experimental Evaluation}
\label{sec:eval}

This section presents our experimental setup (platforms, baselines, and datasets) and our evaluation results.

\subsection{Experimental Setup}
\label{sub:evalsetup}

\subsubsection{Evaluation Platform}

We conduct our experimental evaluation on two HPC systems equipped with NVIDIA Tesla V100 and A100 GPUs\footnote{We note that
	PCIe-A100 holds a marginal 30\% less throughput than SXM4-A100. In this work, we use the SXM4 variant for
	evaluation.}, including ThetaGPU \cite{testbed-thetagpu} at Argonne Leadership Computing Facility and
Longhorn \cite{testbed-longhorn} at Texas Advanced Computing Center.
More details are listed below,
{\small\ttfamily
		\begin{itemize}[leftmargin=10pt]
			\item \textbf{\sffamily ALCF-ThetaGPU}: NVIDIA A100, SXM4 variant, CUDA 11.1
			      \begin{enumerate}[
					      labelsep=10pt,%
					      labelindent=2\parindent,%
					      itemindent=0pt,%
					      leftmargin=*,%
					      listparindent=-\leftmargin
				      ]
				      \item[\sffamily\bfseries DRAM] 40-GB HBM2e at 1555 GB/s {\color{Bv}(1.38$\times$ V100)}
				      \item[\sffamily\bfseries compute] 19.49 FP32 TFLOPS {\color{Bv}(1.73$\times$ V100)}
				      \item[\sffamily\bfseries host] AMD 7532 (32-core), 256 GB
			      \end{enumerate}
			\item \textbf{\sffamily TACC-Longhorn}: NVIDIA V100, SXM2 variant, CUDA 10.2
			      \begin{enumerate}[
					      labelsep=10pt,%
					      labelindent=2\parindent,%
					      itemindent=0pt,%
					      leftmargin=*,%
					      listparindent=-\leftmargin
				      ]
				      \item[\sffamily\bfseries DRAM] 16-GB HBM2 at 900 GB/s
				      \item[\sffamily\bfseries compute] 14.13 FP32 TFLOPS
				      \item[\sffamily\bfseries host] 2 IBM Power9 (40-core), 256 GB
			      \end{enumerate}
		\end{itemize}%
	}

\subsubsection{Baselines}

We compare our {\cuszx} with multiple baselines. Specifically, 1) we compare {\cuszx} with {\cusz} in terms of our
optimized kernels (i.e., Lorenzo construction, Huffman encoding, Lorenzo reconstruction), 2) we compare {\cuszx} on A100
versus on V100, and 3) we compare {\cuszx}'s {Workflow-RLE} with {\cusz}'s {Workflow-Huffman}.

\subsubsection{Test Datasets}
We conduct our evaluation and comparison based on seven typical real-world HPC simulation datasets of each
dimensionality, most of which are from the Scientific Data Reduction Benchmarks suite~\cite{repo-sdrbench}. The datasets
include
	{
		\begin{enumerate}[topsep=0pt]
			\item 1D \texttt{HACC} cosmology particle simulation~\cite{hacc},
			\item 2D \texttt{CESM-ATM} climate simulation~\cite{cesm-atm},
			\item 3D \texttt{Hurricane ISABEL} simulation~\cite{hurricane},
			\item 3D \texttt{Nyx} cosmology simulation~\cite{nyx},
			\item 3D seismic wave \texttt{RTM} data,
			\item 3D hydrodynamics \texttt{Miranda} \cite{miranda} \footnote{It is converted to \texttt{float} from \texttt{double}}, and
			\item 3D \texttt{Quantum Monte Carlo}~\cite{qmcpack}, reinterpreted from 4D.
		\end{enumerate}
	}
They have been widely used in prior
works~\cite{tao2018optimizing,liang2018error,xincluster18,use-case-Franck,liang2019improving} and are good
representatives of production-level simulation datasets. \TAB~\ref{tab:datasets} shows all 128 fields across these
datasets. The data sizes for the seven datasets are 6.3 GB, 2.0 GB, 1.9 GB, 3.0 GB, 1.8 GB, 1.0 GB, and 1.2 GB,
respectively. Note that our evaluated HACC dataset is consistent with real-world scenarios that generate petabytes of
data. For example, according to ~\cite{hacc}, a typical large-scale HACC simulation for cosmological surveys runs on
16,384 nodes, each with 128 million particles, and generates 5 PB over the whole simulation. The simulation contains 100
individual snapshots of roughly 3 GB per node. We evaluate a single snapshot for each dataset instead of all the
snapshots because the compressibility of most of the snapshots usually has strong similarities. Moreover, when the field
is too large to fit in a single GPU's memory, {\cuszx} divides it into blocks and then compresses by block.

\begin{table}[ht]
	\centering
	\resizebox{.8\linewidth}{!}{\footnotesize\centering\ttfamily
\begin{tabular}{@{} >{\bfseries\sffamily}lrr @{}}
                                      & \TABLECAPTION datum size  & \TABLECAPTION \#fields    \\[-.3ex]
    \TABLETITLE datasets              &
    \TABLETITLE dimensions            &
    \TABLETITLE examples(s)                                                                   \\
    \TABLECAPTION cosmology           & \TABLECAPTION 1,071.75 MB & \TABLECAPTION 6 in total  \\[-.6ex]
    HACC                              & 280,953,867               & x, vx                     \\
    \TABLECAPTION climate             & \TABLECAPTION 24.72 MB    & \TABLECAPTION 77 in total \\[-.6ex]
    CESM-ATM                          & 1,800$\times$3,600        & CLDHGH, PHIS              \\
    \TABLECAPTION climate             & \TABLECAPTION 95.37 MB    & \TABLECAPTION 20 in total \\[-.6ex]
    Hurricane                         & 100$\times$500$\times$500 & CLOUDf48, Uf48            \\
    \TABLECAPTION cosmology           & \TABLECAPTION 512 MB      & \TABLECAPTION 6 in total  \\[-.6ex]
    Nyx                               & 512$\times$512$\times$512 & baryon\_desnity           \\
    \TABLECAPTION seismic wave        & \TABLECAPTION 180.72 MB   & \TABLECAPTION 10 in total \\[-.6ex]
    RTM                               & 449$\times$449$\times$235 & snapshot28\{0...9\}0      \\
    \TABLECAPTION hydrodynamics       & \TABLECAPTION 144 MB      & \TABLECAPTION 7 in total  \\[-.6ex]
    Miranda$^\bullet$                 & 256$\times$384$\times$384 & density, pressure         \\
    \TABLECAPTION Quantum Monte Carlo & \TABLECAPTION 601.52 MB   & \TABLECAPTION 2 in total  \\[-.6ex]
    QMCPACK                           & 288x115x69x69             & preconditioned            \\[1ex]
\end{tabular}}
	\caption{Real-world \texttt{float}-type datasets used in the evaluation. $^\bullet\,$\texttt{Miranda} (\texttt{double}-type) is
		converted to \texttt{float}-type for {\cusz}'s support. \texttt{QMCPACK} includes only one field but with two
		representations.}
	\label{tab:datasets}
\end{table}

\subsection{Evaluation on Compression Ratio}

\begin{table}[ht]
	\centering\footnotesize\ttfamily

\begin{adjustbox}{width=.88\linewidth}
	\begin{tabular}{@{} r @{} rrr >{\scriptsize}r r >{\scriptsize}r @{}}
		           &
		\TABLETITLE{\cusz+gzip}
		           &
		\TABLETITLE{\cusz}
		           &
		\multicolumn{2}{c}{\TABLETITLE{ours}}
		           &
		\multicolumn{2}{c@{}}{\TABLETITLE{ours}}
		\\
		           &
		{(\textit{qhg}) ref.}
		           &
		{(\textit{qh}) VLE}
		           &
		{RLE}
		           &
		gain
		           &
		{RLE+VLE}
		           &
		{gain}                                                                     \\
		\cmidrule(lr){4-5}
		\cmidrule(l){6-7}
		AEROD\_v   & 94.27  & 25.06 & 10.46 & -            & 30.33  & 1.21$\times$ \\[-.3ex]
		FLNTC      & 56.95  & 23.66 & 8.87  & -            & 25.35  & 1.07$\times$ \\[-.3ex]
		FLUTC      & 57.06  & 23.66 & 8.91  & -            & 25.46  & 1.08$\times$ \\[-.3ex]
		FSDSC      & 58.30  & 23.88 & 26.10 & 1.09$\times$ & 71.35  & 2.99$\times$ \\[-.3ex]
		FSDTOA     & 430.61 & 26.10 & 43.65 & 1.67$\times$ & 119.17 & 4.57$\times$ \\[-.3ex]
		FSNSC      & 51.73  & 23.44 & 10.11 & -            & 29.46  & 1.26$\times$ \\[-.3ex]
		FSNTC      & 60.35  & 23.88 & 12.33 & -            & 35.50  & 1.49$\times$ \\[-.3ex]
		FSNTOAC    & 111.63 & 25.06 & 12.46 & -            & 35.84  & 1.43$\times$ \\[-.3ex]
		ICEFRAC    & 159.18 & 25.31 & 16.57 & -            & 50.39  & 1.99$\times$ \\[-.3ex]
		LANDFRAC   & 97.15  & 23.66 & 13.98 & -            & 40.50  & 1.71$\times$ \\[-.3ex]
		OCNFRAC    & 89.55  & 23.88 & 11.23 & -            & 32.55  & 1.36$\times$ \\[-.3ex]
		ODV\_bcar1 & 189.28 & 25.83 & 37.28 & 1.44$\times$ & 110.51 & 4.28$\times$ \\[-.3ex]
		ODV\_bcar2 & 197.32 & 25.83 & 30.71 & 1.19$\times$ & 89.98  & 3.48$\times$ \\[-.3ex]
		ODV\_dust1 & 242.89 & 26.10 & 22.91 & -            & 67.72  & 2.59$\times$ \\[-.3ex]
		ODV\_dust2 & 319.55 & 26.37 & 24.02 & -            & 70.98  & 2.69$\times$ \\[-.3ex]
		ODV\_dust3 & 270.50 & 26.10 & 33.29 & 1.28$\times$ & 98.22  & 3.76$\times$ \\[-.3ex]
		ODV\_dust4 & 230.40 & 26.10 & 46.81 & 1.79$\times$ & 139.27 & 5.34$\times$ \\[-.3ex]
		ODV\_ocar1 & 65.81  & 24.11 & 41.17 & 1.71$\times$ & 121.59 & 5.04$\times$ \\[-.3ex]
		ODV\_ocar2 & 64.92  & 24.11 & 33.79 & 1.40$\times$ & 98.63  & 4.09$\times$ \\[-.3ex]
		PHIS       & 98.86  & 25.06 & 9.51  & -            & 28.87  & 1.15$\times$ \\[-.3ex]
		PRECSC     & 176.21 & 25.83 & 19.50 & -            & 58.92  & 2.28$\times$ \\[-.3ex]
		PRECSL     & 142.23 & 25.57 & 15.39 & -            & 45.69  & 1.79$\times$ \\[-.3ex]
		PSL        & 83.13  & 24.34 & 12.43 & -            & 36.32  & 1.49$\times$ \\[-.3ex]
		PS         & 98.59  & 21.09 & 7.45  & -            & 22.27  & 1.06$\times$ \\[-.3ex]
		SNOWHICE   & 144.74 & 25.31 & 15.14 & -            & 45.53  & 1.80$\times$ \\[-.3ex]
		SNOWHLND   & 184.39 & 25.57 & 21.18 & -            & 63.33  & 2.48$\times$ \\[-.3ex]
		SOLIN      & 430.62 & 26.10 & 43.65 & 1.67$\times$ & 119.17 & 4.57$\times$ \\[-.3ex]
		TAUX       & 100.30 & 25.06 & 11.30 & -            & 33.28  & 1.33$\times$ \\[-.3ex]
		TAUY       & 106.55 & 25.31 & 12.40 & -            & 36.45  & 1.44$\times$ \\[-.3ex]
		TREFHT     & 82.50  & 24.58 & 8.75  & -            & 25.12  & 1.02$\times$ \\[-.3ex]
		TREFMXAV   & 87.39  & 24.58 & 9.60  & -            & 27.33  & 1.11$\times$ \\[-.3ex]
		TROP\_P    & 93.78  & 24.82 & 11.19 & -            & 31.40  & 1.27$\times$ \\[-.3ex]
		TROP\_T    & 92.94  & 24.82 & 11.10 & -            & 30.64  & 1.23$\times$ \\[-.3ex]
		TROP\_Z    & 84.81  & 24.58 & 9.48  & -            & 27.07  & 1.10$\times$ \\[-.3ex]
		TSMX       & 64.95  & 23.88 & 8.55  & -            & 24.69  & 1.03$\times$ \\[-.3ex]
	\end{tabular}%
\end{adjustbox}
	\caption{Data fields that {\cuszx} with {Workflow-RLE} has higher compression ratio than {\cusz}
		with {Workflow-Huffman} under $10^{-2}$ error bound. ``gain'' is based on ours
		against (\emph{qh}) VLE from {\cusz}.}

	\label{tab:rle-better-than-vle}
\end{table}

\TAB~\ref{tab:rle-better-than-vle} shows several cases that RLE performs better in compression ratio than \cusz-VLE.
Run-length encoding is implemented using \texttt{thrust::reduce\_by\_key} and achieves 100 GB/s throughput on V100 and
slightly higher throughput on A100. The table demonstrates that RLE may replace the multi-byte VLE in the original
workflow and maintain or achieve higher compression ratios; it can also be used as an additional stage to VLE to get up
to 5.3$\times$ compression ratio improvements over {\cusz} on the tested datasets.

\begin{table}[ht]
	\centering\footnotesize\ttfamily
	\resizebox{.9\linewidth}{!}{
\begin{tabular}{@{} >{\bfseries\sffamily}llrrrrr @{}}
	                       &                   &
	\multicolumn{2}{c}{\TABLETITLE V100 (GB/s)}
	                       &
	\multicolumn{2}{c}{\TABLETITLE A100 (GB/s)}
	                       &
	\TABLETITLE CR
	\\
	                       &                   &
	{\color{gray}Huff}/RLE &
	overall
	                       &
	{\color{gray}Huff}/RLE &
	overall
	                       &
	\\
	\cmidrule(lr){3-4}
	\cmidrule(lr){5-6}
	\cmidrule(l){7-7}
	RTM                    & ours              & 142.4             & 57.8             & 212.6             & 78.0             & \color{Bv}\scriptsize 76.0$\times$   \\
	\#2800                 & \color{gray}\cusz & \color{gray}135.7 & \color{gray}55.1 & \color{gray}233.9 & \color{gray}80.8 & \color{gray}\scriptsize 31.7$\times$ \\
	\cmidrule(lr){3-4}
	\cmidrule(lr){5-6}
	\cmidrule(l){7-7}
	CESM                   & ours              & 104.8             & 47.7             & 162.4             & 57.8             & \color{Bv}\scriptsize 26.1$\times$   \\
	FSDSC                  & \color{gray}\cusz & \color{gray}146.3 & \color{gray}54.8 & \color{gray}146.4 & \color{gray}55.5 & \color{gray}\scriptsize 23.0$\times$ \\
	\cmidrule(lr){3-4}
	\cmidrule(lr){5-6}
	\cmidrule(l){7-7}
	Nyx                    & ours              & 159.1             & 64.1             & 214.5             & 91.2             & \color{Bv}\scriptsize 122.7$\times$  \\
	baryon                 & \color{gray}\cusz & \color{gray}130.8 & \color{gray}58.9 & \color{gray}234.2 & \color{gray}94.8 & \color{gray}\scriptsize 31.0$\times$ \\
\end{tabular}%
}
	\caption{Throughputs (in GB/s) of {\cuszx} (based on RLE) and {\cusz} (based on Huffman coding) on example
		\texttt{RTM}, \texttt{CESM}, and \texttt{Nyx} fields for a demonstration purpose.}
	\label{tab:RLE}
\end{table}

\TAB~\ref{tab:RLE} shows the throughputs of {\cuszx} using the {Workflow-RLE} on the \texttt{RTM},
\texttt{CESM}, and \texttt{Nyx} datasets. It demonstrates that the RLE-based workflow can not only improve the
compression ratio, but also maintain a comparable compression throughput. Thus, {\cuszx} can provide users
flexibility between high compression ratio and performance.

\subsection{Evaluation on Performance and Scalability}
\label{sub:evals}

In this section, we evaluate the compression performance of {\cuszx} and compare it with {\cusz}.

\begin{table}[ht]
	\resizebox{\linewidth}{!}{%
		\centering\footnotesize\ttfamily%
\begin{tabular}{@{} >{\bfseries\sffamily}l *{3}{rr @{\hskip.5em} >{\color{Bv}\scriptsize}r} @{}}
                &
      \multicolumn{3}{c}{\TABLETITLE Lorenzo comp.}
                &
      \multicolumn{3}{c}{\TABLETITLE Huffman Enc.}
                &
      \multicolumn{3}{c}{\TABLETITLE Lorenzo decomp.}                                                         \\
                & {\cusz} &
      \multicolumn{2}{c}{ours}
                & {\cusz} &
      \multicolumn{2}{c}{ours}
                & {\cusz} &
      \multicolumn{2}{c}{ours}
      \\
      \cmidrule(lr){3-4}
      \cmidrule(lr){6-7}
      \cmidrule(l){9-10}
      HACC      & 207.7   & 307.4 & 1.48$\times$ & 54.1 & 58.3  & 1.08$\times$ & 16.8 & 313.1 & 18.64$\times$ \\
      CESM      & 252.1   & 273.9 & 1.09$\times$ & 57.2 & 107.7 & 1.88$\times$ & 58.5 & 254.2 & 4.35$\times$  \\
      Hurricane & 175.8   & 229.9 & 1.31$\times$ & 55.2 & 111.2 & 2.01$\times$ & 43.9 & 218.4 & 4.97$\times$  \\
      Nyx       & 200.2   & 296.0 & 1.48$\times$ & 58.8 & 120.5 & 2.05$\times$ & 29.7 & 238.1 & 8.02$\times$  \\
      QMCPACK   & 189.6   & 298.6 & 1.57$\times$ & 61.0 & 110.8 & 1.82$\times$ & 22.4 & 255.5 & 11.41$\times$ \\
\end{tabular}%
	}
	\caption{Performance comparison of Lorenzo and Huffman encoding kernels in {\cuszx} and {\cusz} on V100. The unit is in GB/s.}
	\label{tab:v100-cmp}
\end{table}

\begin{table*}[ht]
	\centering\footnotesize\ttfamily

\resizebox{\linewidth}{!}{%
    \begin{tabular}{@{} >{\bfseries\sffamily}l  *{7}{@{\hskip.7em}r} r *{7}{@{\hskip.5em}r @{\hskip.5em} >{\scriptsize\color{Bv}}r} @{}}
                                                &
        \multicolumn{7}{c}{\sffamily V100-ours, GB/s}
                                                &       &
        \multicolumn{14}{c}{\sffamily A100-ours, GB/s (and advantage over V100)}
        \\
        \cmidrule[.5pt](r){2-8}\cmidrule[.5pt]{10-23}
        size in MB                              &
        1071.8                                  & 24.7  & 95.4  & 512.0 & 180.7 & 144.0 & 601.5 &       &
        \multicolumn{2}{r}{1071.8}              &
        \multicolumn{2}{r}{24.7}                &
        \multicolumn{2}{r}{95.4}                &
        \multicolumn{2}{r}{512.0}               &
        \multicolumn{2}{r}{180.7}               &
        \multicolumn{2}{r}{144.0}               &
        \multicolumn{2}{r@{}}{601.5}
        \\
                                                &
        \TABLETITLE {HACC}                      &
        \TABLETITLE {CESM}                      &
        \TABLETITLE {Hurr}                      &
        \TABLETITLE {Nyx}                       &
        \TABLETITLE {RTM}                       &
        \TABLETITLE {Mira.}                     &
        \TABLETITLE {QMC}                       &       &
        \multicolumn{2}{r}{\TABLETITLE HACC}    &
        \multicolumn{2}{r}{\TABLETITLE CESM}    &
        \multicolumn{2}{r}{\TABLETITLE Hurr}    &
        \multicolumn{2}{r}{\TABLETITLE Nyx}     &
        \multicolumn{2}{r}{\TABLETITLE RTM}     &
        \multicolumn{2}{r}{\TABLETITLE Miranda} &
        \multicolumn{2}{r@{}}{\TABLETITLE QMC}
        \\
        \cmidrule(r){10-11}
        \cmidrule(r){12-13}
        \cmidrule(r){14-15}
        \cmidrule(r){16-17}
        \cmidrule(r){18-19}
        \cmidrule(r){20-21}
        \cmidrule{22-23}
        Lorenzo construct                       & 328.3 & 273.9 & 199.0 & 296.0 & 193.1 & 289.3 & 298.6 &   & 501.1 & 1.53$\times$ & 466.8 & 1.70$\times$ & 429.0 & 2.16$\times$ & 481.3 & 1.63$\times$ & 422.7 & 2.19$\times$ & 480.7  & 1.66$\times$ & 492.9 & 1.65$\times$ \\
        gather outlier                          & 221.4 & 160.6 & 251.1 & 238.0 & 249.7 & 228.6 & 261.2 &   & 324.8 & 1.47$\times$ & 151.4 & 0.94$\times$ & 284.2 & 1.13$\times$ & 334.9 & 1.41$\times$ & 221.6 & 0.89$\times$ & 336.0  & 1.47$\times$ & 266.2 & 1.02$\times$ \\
        histogram                               & 565.9 & 356.5 & 438.4 & 372.4 & 573.6 & 489.8 & 724.3 &   & 923.5 & 1.63$\times$ & 409.8 & 1.15$\times$ & 681.2 & 1.55$\times$ & 870.2 & 2.34$\times$ & 793.9 & 1.38$\times$ & 714.9  & 1.46$\times$ & 569.7 & 0.79$\times$ \\
        Huffman encode                          & 58.3  & 107.7 & 111.2 & 120.5 & 123.2 & 161.1 & 110.8 &   & 174.6 & 2.99$\times$ & 121.6 & 1.13$\times$ & 206.0 & 1.85$\times$ & 217.2 & 1.80$\times$ & 202.2 & 1.64$\times$ & 201.6  & 1.25$\times$ & 198.4 & 1.79$\times$ \\
        overall, compress                       & 42.1  & 44.8  & 49.3  & 53.9  & 52.5  & 62.2  & 56.9  &   & 84.1  & 2.00$\times$ & 51.5  & 1.15$\times$ & 82.2  & 1.67$\times$ & 92.4  & 1.72$\times$ & 76.4  & 1.46$\times$ & 87.6   & 1.41$\times$ & 79.5  & 1.40$\times$ \\
                                                &       &       &       &       &       &       &       &   &       &              &       &              &       &              &       &              &       &              &        &              &       &              \\
        Huffman decode                          & 42.1  & 37.9  & 45.8  & 66.8  & 48.9  & 42.7  & 44.6  &   & 48.5  & 1.15$\times$ & 26.6  & 0.70$\times$ & 51.8  & 1.13$\times$ & 91.2  & 1.37$\times$ & 56.0  & 1.15$\times$ & 50.1   & 1.17$\times$ & 49.0  & 1.10$\times$ \\
        scatter outlier                         & 225.0 & 334.8 & 628.1 & 359.7 & 440.2 & 679.1 & 347.1 &   & 658.4 & 2.93$\times$ & 630.2 & 1.88$\times$ & 918.3 & 1.46$\times$ & 797.4 & 2.22$\times$ & 906.6 & 2.06$\times$ & 1066.8 & 1.57$\times$ & 782.8 & 2.26$\times$ \\
        Lorenzo reconstruct                     & 308.7 & 267.0 & 200.1 & 251.7 & 201.3 & 245.3 & 255.5 &   & 504.4 & 1.63$\times$ & 495.3 & 1.86$\times$ & 345.5 & 1.73$\times$ & 398.6 & 1.58$\times$ & 335.6 & 1.67$\times$ & 386.9  & 1.58$\times$ & 384.0 & 1.50$\times$ \\
        overall, decompress                     & 31.8  & 30.2  & 35.2  & 46.0  & 36.1  & 34.5  & 34.2  &   & 41.4  & 1.30$\times$ & 24.3  & 0.80$\times$ & 43.0  & 1.22$\times$ & 67.9  & 1.47$\times$ & 45.6  & 1.26$\times$ & 42.6   & 1.23$\times$ & 41.2  & 1.20$\times$ \\
    \end{tabular}%
}%
	\caption{Evaluation of {\cuszx} using default compression workflow (Lorenzo and multi-byte VLE) with relative error
		bound of $10^{-4}$ on V100 and A100: breakdown throughput of compression subprocedures.}
	\vspace{-2mm}
	\label{tab:eval-default}
\end{table*}

\subsubsection{Evaluation on Optimized Kernels}
We first evaluate the performance of the majorly changed kernels in {\cuszx} and {\cusz} on V100, as shown in
\TAB~\ref{tab:v100-cmp}. The baseline is from the evaluation results shown in the {\cusz} paper \cite{cusz2020}. The
table illustrates that the performance improvements of {\cuszx}'s Lorenzo construction kernels are 1.48$\times$ for 1D
data, 1.09$\times$ for 2D data, and 1.45$\times$ for 3D data on average over {\cusz}. Moreover, we increase the lowest
throughput from 175.8 GB/s to 229.9 GB/s (+30.7\%) on the tested datasets.

For Huffman encoding kernel, despite it being more latency-bound, it also suffers from non-coalescing store. Because of the
variable-length encoding and bit operation spanning multiple bytes, it is impossible to make threads work in a
synchronized manner (otherwise, a high synchronization overhead would be imposed). Our optimization can decrease the
number of DRAM store transactions to be inversely proportional to the compression ratio. In particular, we perform a
DRAM store only when a new data unit needs to be written back, which helps us achieve 1.08$\times$ to 2.05$\times$
performance gain.

The table also shows that by using the fine-grained $N$D partial-sum computation, {\cuszx}'s Lorenzo reconstruction
kernel exhibits up to 18.4$\times$ performance improvements over {\cusz}'s coarse-grained kernel. The 2D and 3D kernels
also exhibit 4.5$\times$ and 4.6$\times$ to 7.1$\times$ performance improvements, respectively. In addition,
\TAB~\ref{tab:eval-default} shows the speedup of our optimized kernels on A100 compared to on V100.

\subsubsection{Evaluation on Default Compression Workflow}
The rest of \TAB~\ref{tab:eval-default} shows the evaluation of {\cuszx} based on the default compression
workflow on both V100 and A100 with the relative error bound of $10^{-4}$ (with PSNRs higher than 85 dB). It illustrates
the performance of our optimized compression and decompression kernels. We note that the performance improvements of
the histogram kernel and the gather-outlier kernel are relatively lower than those of other compression kernels (the
performance is even degraded on some datasets such as CESM and RTM) from using A100 to using V100. The degradation may be because
each field of \texttt{CESM-ATM} and \texttt{RTM} is fairly small (24.7 MB and 180 MB, respectively), such that the
histogram kernel (using the algorithm from \cite{gomez2013optimized}) and the gather-outlier kernel (using the dense-to-sparse kernel from
cuSPARSE) on A100 do not maintain the same work efficiency as on V100.
But note that these two kernels would not be bottlenecks
\footnote{The Huffman encoding kernel is the main bottleneck in the compression workflow compared to other compression kernels.} for a relatively large dataset (e.g., hundreds of MBs per field), which is more
common in practice (e.g., \texttt{HACC} and \texttt{Nyx}).

In addition, we observe that {\cuszx}'s kernels with different parallelism have different scalabilities. Specifically, the sparsity-related operation in compression can be enhanced significantly by using A100 GPU, but other compression kernels such as Huffman encoding are scaled up marginally.
As a result, the overall improvement of compression
performance is limited when changing from using V100 to using A100. Similarly, for decompression, although
the outlier scatter operation scales naturally and shows a speedup of higher than $2\times$, the multi-byte Huffman
decoding exhibits a stagnation in scaling up, resulting in a marginal improvement of the overall decompression
performance.

\section{Related Work}
\label{sec:related}

Compression for scientific datasets has been studied for years to reduce the storage burden and I/O overhead.
Scientific data compression techniques fall into two classes: lossless compression and lossy
compression. The former includes the generic lossless compressors such as Zlib~\cite{gzip} and Zstd \cite{zstd}, as well
as the specific algorithm designed for floating-point values such as FPZIP~\cite{lindstrom2006fast} and FPC~\cite{fpc}.
The lossless compressors, however, all suffer from very low compression ratios (generally 2:1 or even lower
\cite{son2014data}) because of the somewhat random ending mantissa bits in the floating-point representation.

Lossy compressors have been studied for decades. The traditional lossy compressors (such as JPEG~\cite{jpeg} and
JPEG2000 \cite{jpeg2000}) are designed for 2D images, which are not suitable for scientific datasets. The key reason is
that the traditional lossy compressors focus on the visual quality while scientific applications care more about the
post hoc analysis results beyond the simple visualization purpose.

To address the above-mentioned gap, error-bounded lossy compressors \cite{zfp,sz16,sz17,fpzip,ssem} have been proposed
for years. Based on their design principle, they can be split into two categories - prediction-based model
\cite{sz16,sz17,liang2018error} and transform-based model \cite{zfp,ssem,vapor}. The typical example in the former
category is SZ, which supports different categories of errors to control data distortion, such as absolute error bound,
relative error bound, and peak signal-to-noise ratio (PSNR). The typical example in the latter category is ZFP, which
supports absolute error bound and precision mode%
\footnote{In the precision mode, users can use an integer number to control the data distortion. Higher precision in value means lower data distortion.}.

However, all the above existing lossless and lossy compressors cannot run on GPUs directly. Recently, both the
SZ team and the ZFP team released their CUDA versions, called \cusz~\cite{cusz2020} and cuZFP~\cite{cuZFP}, respectively.
Both versions provide much higher throughputs for compression and decompression compared with their CPU versions.
Compared with \cusz, cuZFP provides slightly higher compression throughput, but it only supports fixed-rate mode,  significantly limiting its adoption in practice.
In comparison with the two existing GPU-supported compressors, our designed new compression method is aware of the compressibility of the datasets, such that it can adopt the run-length
encoding (RLE) method to significantly improve the compression ratios when needed.

\section{Conclusion and Future Work}
\label{sec:conclusion}

In this work, we propose {\cuszx}, a compressibility-aware GPU-based lossy compressor for NVIDIA GPU architectures,
which can effectively improve the compression throughput over \cusz.
Specifically, we propose an efficient compression method to adaptively perform run-length encoding and/or Huffman encoding by considering data smoothness to improve the compression ratio over {\cusz}.
We prove that the Lorenzo reconstruction in decompression is equivalent to a multidimensional partial-sum computation and develop an efficient fine-grained Lorenzo reconstruction algorithm on GPUs.
Moreover, we carefully optimize \cusz compression kernels by leveraging different techniques for CUDA architectures.
Finally, we evaluate {\cuszx} using seven real-world HPC application datasets on the
most advanced GPUs (V100 and A100) and compare with {\cusz}, the GPU-centric error-bounded lossy compressor. Experiments show
that our {\cuszx} improves {\cusz}'s decompression kernel throughput by up to {1.6$\times$} with the same compression
quality on state-of-the-art GPUs, including A100.

In the future, we plan to optimize the performance of decompression further, implement other data prediction methods
such as linear-regression-based predictors and evaluate the performance improvements of parallel I/O with {\cusz}.

\section*{Acknowledgments}
This research was supported by the Exascale Computing Project (ECP), Project Number: 17-SC-20-SC, a collaborative
effort of two DOE organizations---the Office of Science and the National Nuclear Security Administration, responsible
for the planning and preparation of a capable exascale ecosystem, including software, applications, hardware, advanced
system engineering, and early testbed platforms, to support the nation's exascale computing imperative. The material was
supported by the U.S. Department of Energy, Office of Science, under contract DE-AC02-06CH11357. This work was also
supported by the National Science Foundation under Grants OAC-2042084, OAC-2034169, OAC-2003709, and CCF-1619253.

\newpage
\renewcommand*{\bibfont}{\footnotesize}
\printbibliography[]

\end{document}